\documentclass[12pt,twoside,spanish]{article}
\usepackage{amsmath}
\usepackage{amssymb,amsfonts,amsthm}
\usepackage[pdftex]{graphicx} 
\usepackage{url} 
\usepackage{bm} 
\usepackage[pdftex,breaklinks=true]{hyperref} 
\usepackage{slashed}
\usepackage[spanish]{babel}
\usepackage[usenames,dvipsnames]{xcolor}
\usepackage[utf8]{inputenc}
\usepackage{fancyhdr}

\usepackage[sort&compress,comma]{natbib}
\bibpunct{[}{]}{,}{n}{}{,}

\usepackage[light]{antpolt}
\usepackage[T1]{fontenc}

\hypersetup{%
  pdftitle    = {Agujeros negros cu\'anticos en la teoría de cuerdas tipo-IIA},
  pdfkeywords = {Teor\'ia de cuerdas, agujeros negros, Calabi Yau, Compactificaci\'on, Correcciones cu\'anticas, Conjetura de no-pelo},
  pdfauthor   = {Pablo Bueno},
  plainpages  = true,
  colorlinks  = true,
  citecolor   = Red	,
  urlcolor    = Green,
  linkcolor   = Blue
}

\def\-{\hphantom{-}}

\def\s2{\frac{1}{\sqrt2}}

\def\beq{\begin{equation}}
\def\eeq{\end{equation}}
\def\beqa{\begin{eqnarray}}
\def\eeqa{\end{eqnarray}}

\def\Dsl{\,\raise.15ex\hbox{/}\mkern-13.5mu D} 

\def\be{\begin{equation}}
\def\ee{\end{equation}}
\def\bea{\begin{eqnarray}}
\def\eea{\end{eqnarray}}

\begin{document}

\makeatletter
\@addtoreset{equation}{section}
\makeatother
\renewcommand{\theequation}{\thesection.\arabic{equation}}

\pagestyle{headings}
\pagestyle{empty}

\vspace*{-1.3cm}
\vspace{-1cm}
\newcommand{\HRule}{\rule{\linewidth}{1mm}}
\setlength{\parindent}{1cm}
\setlength{\parskip}{1mm}
\noindent
\HRule
\begin{center}
\Huge{\textbf{Agujeros negros cuánticos en la teoría de cuerdas tipo-IIA}}
 \\[5mm]
\end{center}
\HRule

\vspace{1.2cm}

\begin{center}

	 \large{
	   Memoria de trabajo presentada por \textbf{Pablo Bueno Gómez} \\
                  para optar al t\'itulo de \textbf{M\'aster en F\'isica Te\'orica} por la\\
                  \textbf{Universidad Aut\'onoma de Madrid}\\
\vspace{0.8cm}
Trabajo dirigido por \textbf{Prof. Don Tomás Ortín Miguel} \\
Profesor de Investigación en el Instituto de Física Teórica UAM/CSIC, Madrid. }

\vspace{0.4cm}

\end{center}
\vspace{.5cm}

\begin{figure}[ht]
\centering
\begin{tabular}{lr}

\includegraphics[scale=0.12]{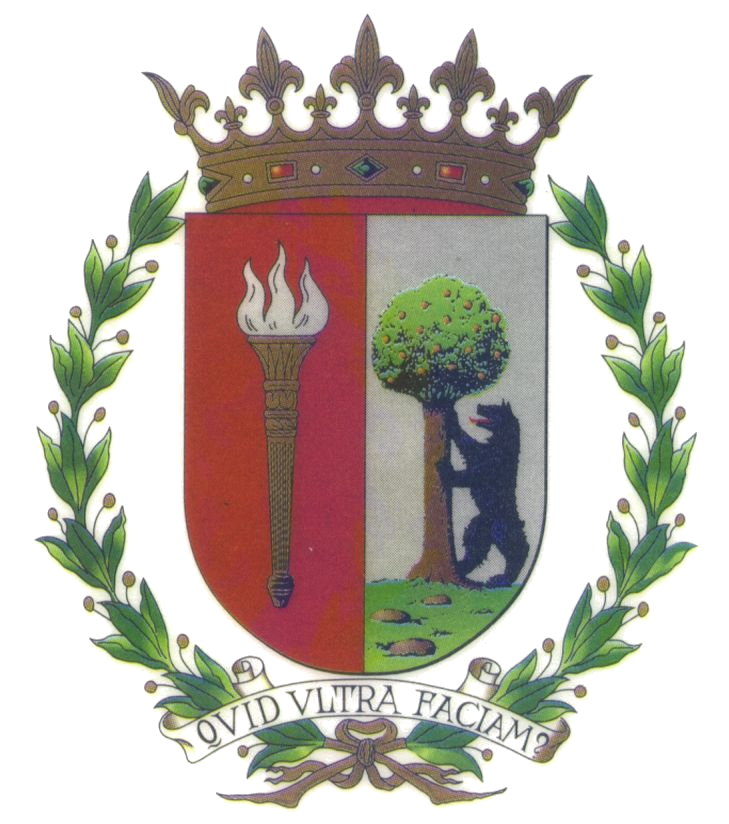}\qquad\qquad\qquad
&
\qquad\qquad\qquad\includegraphics[scale=0.54]{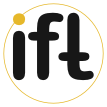}

\end{tabular}
\end{figure}

\vspace{0.2cm}

\begin{center}
{\Large {Departamento de F\'isica Te\'orica\\ Universidad Aut\'onoma de Madrid }}\\
\vspace{0.2cm}
{\Large {Instituto de F\'isica Te\'orica\\ \vspace{0.1cm} UAM/CSIC}}\\

\end{center}

\vspace{0.2cm}
\begin{center}
2 de Octubre de 2013
\end{center}

\newpage
\thispagestyle{empty}
\phantom{asdfaf}

\newpage
\thispagestyle{empty}

\newpage

\thispagestyle{empty}
\begin{center}
\textbf{Resumen}
\end{center}
\begin{centering}
En el contexto de la teoría de cuerdas tipo-IIA compactificada en una variedad de Calabi-Yau (CY) a un espaciotiempo de cuatro dimensiones, estudiamos el efecto de la introducción de correcciones perturbativas y no perturbativas (en $\alpha^{\prime}$) en la teoría de supergravedad efectiva resultante sobre el espacio de soluciones de tipo agujero negro. Considerando el límite de gran volumen del CY, en el que las correcciones no perturbativas están exponencialmente suprimidas, comenzamos por definir una nueva familia de soluciones que resultan ser \textit{genuinamente} cuánticas, en el sentido de que no solo las soluciones carecen de límite clásico bien definido, sino que la truncación considerada se vuelve inconsistente en tal régimen. A continuación construimos la primera solución de agujero negro no extremo con escalares no constantes en presencia de correcciones perturbativas. Seguidamente, restringiéndonos al caso de variedades de Calabi-Yau \textit{auto-especulares}, estudiamos el caso en el que la contribución subdominante es de origen no perturbativo, obteniendo la primera familia de soluciones explícitas supersimétricas de este tipo. De forma inesperada, tal familia resulta venir dada en términos de una función multivaluada, lo que da lugar a una posible violación de la conjetura de no-pelo. Argumentamos de qué manera tal posibilidad no es posible en teoría de cuerdas, y cómo permanece abierta en el contexto de la supergravedad \textit{no gaugeada} en cuatro dimensiones.
\end{centering}

\vspace{1cm}

Esta memoria está basada en:
\begin{itemize}
\item \cite{Bueno:2012jc} P.~B., Rhys~Davies y C.~S.~Shahbazi,``Quantum black holes in Type-IIA String Theory,'' JHEP {\bf 1301} (2013) 089, arXiv:1210.2817 [hep-th]]
\item \cite{Bueno:2013psa} P.~B. y C.~S.~Shahbazi, ``Non-perturbative black holes in Type-IIA String Theory vs. the No-Hair conjecture,'' a aparecer en: Classical and Quantum Gravity, arXiv:1304.8079 [hep-th]
\end{itemize}

\newpage

\thispagestyle{empty}
\phantom{asdfaf}


\newpage
\thispagestyle{empty}
\tableofcontents

\newpage

\pagestyle{fancy}
\fancyhead{}
\fancyfoot{}

\fancyhead[LE,RO] { \slshape \leftmark}
\fancyfoot[C]{\thepage}
\renewcommand{\headrulewidth}{0.3pt}
\renewcommand{\footrulewidth}{0pt}

\setcounter{page}{1}

\section{Introducción}
La física fundamental contemporánea se asienta sobre dos pilares sostenidos por una evidencia experimental sin precedentes en la historia de la ciencia, tanto por la cantidad y diversidad de los experimentos, como por su precisión (véanse, por ejemplo, \cite{Turyshev:2008dr,Aspect:1982fx}). Nos referimos, obviamente, a la teoría de la relatividad general (RG), que describe la interacción gravitatoria; y a la mecánica cuántica (MC), que ha alcanzado su máxima expresión a través de las teorías cuánticas de campos, en las que se basa el increíblemente exitoso modelo estándar (ME) de la física de partículas, que acomoda todas las interacciones fundamentales conocidas, a excepción precisamente de la gravedad.

Ya desde mediados del siglo XX, el problema central de la física teórica ha consistido en la búsqueda de un marco unificado que permita compatibilizar ambas teorías\footnote{Existen, naturalmente, otros problemas sin resolver que se enmarcan en el contexto de cada una de las teorías (RG y MC), por ejemplo: el problema de la materia oscura, el de la energía oscura y la constante cosmológica, el problema de la jerarquía y el fine-tuning de la masa del Higgs, la masa de los neutrinos y las anomalías entre experimentos, el Strong CP problem, el momento magnético anómalo del muón, etcétera.} y, finalmente, ofrecer una descripción cuántica unificada de todas las interacciones fundamentales, incluyendo la gravedad.

Una de la propuestas más prometedoras hasta la fecha es la teoría de cuerdas \cite{Green:1987sp,Green:1987mn,Becker:2007zj,Polchinski:1998rq,Polchinski:1998rr,Ortin:2004ms}, que es de hecho una teoría cuántica de la gravedad presumiblemente internamente consistente. A día de hoy no existe evidencia experimental alguna de que tal teoría sea correcta, y no parece posible que con la tecnología disponible en la actualidad (o en un futuro más o menos próximo), un experimento susceptible de ponerla a prueba pueda ser concebido. Sin embargo, la teoría sí ha superado con éxito varias de las pruebas de consistencia que toda candidata a teoría del todo que se precie (si es que alguna ha de haber) habría de satisfacer \cite{Green:1984sg,Green:1987sp,Green:1987mn,Becker:2007zj,Polchinski:1998rq,Polchinski:1998rr}.

Una de esas pruebas consiste en el acuerdo entre el valor de la entropía macroscópica de Bekenstein-Hawking asociada a cualquier solución de tipo agujero negro embebida en la teoría\footnote{Usualmente encontrada como solución a la teoría de supergravedad que corresponda al límite de bajas energías de la teoría de supercuerdas en cuestión, posiblemente compactificada a un espaciotiempo de dimensión menor que 10.}\footnote{Y que, en el caso clásico viene dada por la ley de área de Bekenstein-Hawking:  S = A/4. \cite{Bekenstein:1973ur,Hawking:1971tu,Bardeen:1973gs} (Véase la sección \ref{BHSugra}).}, y el de la entropía microscópica, obtenido a través del contaje del número de microestados accesibles a los grados de libertad fundamentales de la teoría \cite{Strominger:1996sh}. Tal acuerdo ha sido comprobado con éxito para ciertos tipos de soluciones extremas, tanto al orden más bajo en curvatura, como considerando correcciones de orden superior \cite{LopesCardoso:1998wt,LopesCardoso:1999ur,LopesCardoso:2000qm,Mohaupt:2000mj}, las cuales modifican la forma de la ley de área para la entropía.

El límite de baja energía de la teoría tipo-IIA, que es una de las cinco teorías de supercuerdas (principales), compactificada a 4 dimensiones en un CY viene dada por una teoría de supergravedad (al igual que ocurre para las otras cuatro teorías) \cite{Ortin:2004ms,Freedman:2012zz} correspondiente al nivel árbol en la constante de acoplo de la cuerda $g_s$, con la aparición de correcciones de orden más alto en curvatura ya a 1 loop \cite{Mohaupt:2000mj}. Existe, sin embargo, otro tipo de correcciones cuerdosas (que corrigen el comportamiento de partícula puntual propio de las teorías de campos), que no modifican el lagrangiano efectivo de la tipo-IIA con términos más altos en la curvatura, sino que modifican los acoplamientos y la geometría de la variedad escalar en la acción de la teoría de supergravedad correspondiente. Estas son (algunas de) las correcciones en $\alpha^{\prime}$\footnote{Perturbativas (world-sheet loops) y no perturbativas (instantones de la world-sheet).}, usualmente llamadas cuánticas en la literatura\footnote{Al menos en el contexto de las compactificaciones de la tipo-IIA en variedades CY.} \cite{Candelas:1990qd,Candelas:1990rm,Hosono:1994ax,Hosono:1994av}. Dado que no incluyen términos de orden más alto en curvatura, estas correcciones no modifican la ley de área\footnote{Obviamente, estas correcciones sí modifican el área del agujero negro en cuestión, en tanto en cuanto la estructura de la solución cambia cuando se tienen en cuenta.}.

Ocurre que los efectos de este tipo de correcciones sobre las propiedades de las soluciones de tipo agujero negro de las distintas teorías de cuerdas (y en particular de la tipo-IIA) no han sido apenas estudiadas (en \cite{Behrndt:1997ei,Gaida:1997id,Behrndt:1997gs,Gaida:1998pz,Galli:2012pt} pueden encontrarse algunas excepciones), quedando en entredicho la posible igualdad entre las entropías macroscópica y microscópica de todo el (inmenso y casi) inexplorado espacio de soluciones de tipo agujero negro en presencia de correcciones cuánticas.

Por otro lado, la obtención de soluciones no extremas en teoría de cuerdas, y el estudio de sus propiedades físicas (incluyendo una comparación de las correspondientes entropías microscópica y macroscópica, que hoy en día parece completamente fuera de alcance) ha sido y continúa siendo un área de investigación activa y abierta, dada la complejidad técnica de la tarea y el interés de una exploración completa del espacio de soluciones de tipo agujero negro de las distintas teorías de cuerdas lejos de los regímenes supersimétrico o extremo (este incluyendo al anterior)\cite{Ortin:2012mt,Galli:2011zz,Mohaupt:2012tu,Mohaupt:2010fk,Toldo:2012ec}.

En este trabajo vamos a hacer uso del formalismo H-FGK, desarrollado en \cite{Mohaupt:2010fk,Mohaupt:2011aa,Meessen:2011bd,Meessen:2011aa,Meessen:2012su,Galli:2012ji} en el contexto de la supergravedades $\mathcal{N} = 2$, $d = 4, 5$ con el objetivo de construir nuevas soluciones de tipo agujero negro estáticas y esféricamente simétricas, para abordar ambas cuestiones: la exploración del espacio de agujeros negros no extremos en la tipo-IIA, y el estudio de los efectos de las mencionadas correcciones cuánticas sobre las propiedades físicas de tales soluciones. Como veremos, este estudio dará lugar a interesantes e inesperados resultados.

En la sección \ref{CG} introducimos las herramientas de la geometría diferencial compleja necesarias para la formulación de la teoría de supergravedad $\mathcal{N}=2,$ $d=4$, que llevamos a cabo en la sección siguiente (\ref{N=2}). Para ello, comenzamos por repasar brevemente la estructura de las teorías supersimétricas y de supergravedad, así como la existencia de una dualidad eléctrico-magnética generalizada en el sector bosónico de las teorías de supergravedad en cuatro dimensiones. Finalmente resumimos la formulación geométrica del sector bosónico de la teoría de supergravedad $\mathcal{N}=2,$ $d=4$ \textit{no gaugeada}.

En la \ref{IIA} repasamos brevemente la formulación de la teoría de cuerdas tipo-IIA compactificada en un CY de dimensión compleja 3, mientras que en la sección \ref{BHSugra} repasamos algunos aspectos fundamentales de la termodinámica de agujeros negros, así como el papel de estos objetos en las teorías de supergravedad.

En las secciones \ref{QBH} y \ref{NPBH} se presentan los resultados originales de la tesis. En la primera de ellas (basada en \cite{Bueno:2012jc}), comenzamos por explicar en qué límite pueden considerarse subdominantes las correcciones cuánticas perturbativas y no perturbativas que vamos a considerar, así como la truncación consistente que utilizaremos durante todo el trabajo. A continuación, considerando solo la contribución cuántica perturbativa (que es la dominante en el límite de gran volumen del CY), construimos una nueva clase de agujeros negros que resultan ser esencialmente cuánticos, puesto que las soluciones carecen de límite clásico bien definido, y la truncación considerada se vuelve inconsistente en tal régimen. Estudiamos también cómo la existencia de tales soluciones impone una condición sobre la topología de la variedad de CY en cuestión, y presentamos nuevas soluciones no extremas para dos modelos definidos por dos variedades de compactificación con números de Hodge pequeños, explícitamente construidas en \cite{Bueno:2012jc}. Seguidamente, construimos la primera solución de agujero negro no extremo con escalares no constantes en presencia de correcciones cuánticas perturbativas existente en la literatura\footnote{Poco después de \cite{Bueno:2012jc}, apareció el artículo \cite{Galli:2012pt}, en el que también se construían nuevas soluciones de este tipo.}.

En la siguiente sección (basada en \cite{Bueno:2013psa}), restringiéndonos al caso de CY auto-especulares, estudiamos el caso en el que la contribución subdominante es de origen no perturbativo, y obtenemos una nueva familia de soluciones explícitas supersimétricas de este tipo. Sorprendentemente, tal familia resulta venir dada en términos de una función multivaluada, lo que da lugar a una posible violación de la conjetura de no-pelo. Finalmente, analizamos las propiedades y consecuencias de nuestras nuevas soluciones y argumentamos que la posible violación de la conjetura de no-pelo no resulta posible en teoría de cuerdas. Esta permanece no obstante abierta en el contexto de la supergravedad \textit{no gaugeada} en cuatro dimensiones.

\newpage
\section{Geometría diferencial compleja}
\label{CG}
En esta sección\footnote{Existen numerosísimas introducciones a la geometría diferencial compleja en la literatura. Por ejemplo \cite{Napoli:2003,nakahara2003geometry,Ortin:2004ms,Shahbazi:2013ksa}} vamos a repasar una serie de resultados de la geometría diferencial compleja, intímamente relacionados con la estructura matemática de todas las teorías supersiméticas, y en particular de la supergravedad $\mathcal{N}=2,~d=4$.
\subsection{Variedades casi complejas}
\textit{Definición 1.}\\
Sea $\mathcal{M}$ una variedad diferenciable\footnote{Es decir, un espacio topológico Hausdorff, completamente separable equipado con una estructura diferenciable (y por tanto, como consecuencia del \textit{teorema de metrización de Urysohn}, también paracompacto y metrizable).}. El par $(\mathcal{M},\mathcal{G})$ es una \textit{variedad riemanniana} si $\mathcal{G}$ es una \textit{métrica riemanniana}, es decir, un tensor suave 2-covariante ($\mathcal{G}_p\in T_p^*\mathcal{M}\otimes T_p^*\mathcal{M}$ en cada punto $p\in \mathcal{M}$) definido globalmente sobre $\mathcal{M}$, simétrico, no degenerado y definido positivo $\forall p \in  \mathcal{M}$.\\

\textit{Definición 2.}\\
Sea $\mathcal{M}$ una variedad diferenciable. El par $(\mathcal{M},\mathcal{\omega})$ es una \textit{variedad simpléctica} si $\mathcal{\omega}$ es una 2-forma suave (antisimétrica: $\omega_p \in T_p^*\mathcal{M}\wedge  T_p^*\mathcal{M}$ en cada punto $p\in \mathcal{M}$) no degenerada y globalmente definida sobre $\mathcal{M}$.\\

\textit{Definición 3.}\\
Sea $\mathcal{M}$ una variedad diferenciable. El par $(\mathcal{M},\mathcal{J})$ es una \textit{variedad casi compleja} si $\exists$ un campo tensorial $\mathcal{J}_p(\in T_p\mathcal{M} \otimes T^*_p\mathcal{M}):T_p\mathcal{M}\rightarrow T_p\mathcal{M}$ tal que $\forall p\in \mathcal{M}$, $\mathcal{J}_p^2=-1$. En tal caso, $\mathcal{J}$ se dice una \textit{estructura casi compleja}. Sus propiedades son:

\begin{itemize}
\item Puesto que $\mathcal{J}_p^2=-1$, $\mathcal{J}_p$ tiene valores propios $\pm i$. Si hay $n$ valores propios $+i$, habrá otros tantos $-i$, de forma que $\mathcal{J}_p$ actuará en coordenadas como una matriz $2n\times 2n$. Como consecuencia, solamente variedades de dimensionalidad par pueden ser dotadas de una estructura casi compleja.
\item La estructura casi compleja permite dividir el espacio tangente complexificado\footnote{$T_p\mathcal{M}^{\mathbb{C}}\equiv T_p \mathcal{M} \otimes \mathbb{C}$} en dos subespacios vectoriales (\textit{holomorfo} y \textit{antiholomorfo} respectivamente) disjuntos
    \begin{equation}
    T_p\mathcal{M}^{\mathbb{C}}=T_p\mathcal{M}^+\oplus T_p\mathcal{M}^-\, ,
    \end{equation}
    donde $\oplus$ es la \textit{suma de Withney} y
    \begin{equation}
    T_p\mathcal{M}^{\pm}=\left\{Z\in T_p\mathcal{M}^{\mathbb{C}} \slash \mathcal{J}_p Z=\pm i Z  \right\}\, .
    \end{equation}
De esta forma, cualquier $Z\in T_p\mathcal{M}^{\mathbb{C}}$ puede reescribirse como $Z=Z^++ Z^-$ con $Z^{\pm}\equiv \mathcal{P}^{\pm}Z\in T_p\mathcal{M}^{\pm}$, donde hemos definido los proyectores $\mathcal{P}^{\pm}:T_p\mathcal{M}^{\mathbb{C}}\rightarrow T_p\mathcal{M}^{\pm}$, $\mathcal{P}^{\pm}=\frac{1}{2}(\mathbb{I}\mp i \mathcal{J}_p)$, de manera que se tiene

\begin{equation}
\mathcal{J}_p(\mathcal{P}^{\pm}Z)=\mathcal{J}_p Z^{\pm}=\pm i (\mathcal{P}^{\pm}Z)=\pm i Z^{\pm}\, .
\end{equation}

\end{itemize}

\subsection{Variedades complejas}
Si la acción de $\mathcal{J}$ es independiente de la carta, es decir, si la estructura casi compleja tiene la misma expresión en todos los puntos de la variedad, la estructura compleja de $\mathcal{M}$ queda completamente determinada, y $\mathcal{J}$ se dirá \textit{integrable}. De forma más precisa tenemos:\\

\textit{Definición 4.}\\
Sea $\mathcal{J}$ una estructura casi compleja. $\mathcal{J}$ se dice integrable $\iff$ $\mathcal{N}(u,v)=0~\forall~u,v\in T_p\mathcal{M}$. $\mathcal{N}(.,.)$ es el llamado \textit{tensor de Nijenhuis} (que actúa como una especie de torsión para $\mathcal{J}$),
\begin{equation}
\mathcal{N}(u,v)\equiv \left[\mathcal{J}u,\mathcal{J}v\right] - \mathcal{J}\left[u,\mathcal{J}v\right]-\mathcal{J}\left[\mathcal{J}u,v\right]-\left[u,v\right]\, ,
\end{equation}
e intuitivamente parametriza la imposibilidad de definir cambios de coordenadas holomorfos en $\mathcal{M}$. Cuando $\mathcal{N}(.,.)$ se anula, $\mathcal{M}$ puede cubrirse con un atlas holomorfo, y se dice una variedad compleja:\\

\textit{Definición 5.}\\
Una variedad casi compleja $(\mathcal{M} ,\mathcal{J})$ es una \textit{variedad compleja} si en la intersección de dos cartas cualesquiera $(U_i,\phi_i)$ y $(U_j,\phi_j)$, el mapa $\psi_{ij}\equiv \phi_j\phi^{-1}_i: \phi_i(U_i\cap U_j)\rightarrow \phi_j(U_i\cap U_j)$ es holomorfo.\\ Consideremos ahora dos cartas $(U,\phi)$, $(V,\psi)$ con intersección no vacía sobre una variedad compleja $(\mathcal{M} ,\mathcal{J})$. Podemos definir:
\begin{equation}
\phi(p)\equiv z^{\mu}\equiv  x^{\mu}+i y^{\mu},~~ \psi(p)\equiv w^{\mu}\equiv  u^{\mu}+i v^{\mu},~~1\leq \mu, \nu \leq n\, ,
\end{equation}
de forma que $z^{\mu}(w^{\nu})$ es holomorfa, y por tanto satisface las ecuaciones de Cauchy-Riemann
\begin{equation}
\frac{\partial u^{\nu}}{\partial x^{\mu}}=\frac{\partial v^{\nu}}{\partial y^{\mu}},~~\frac{\partial u^{\nu}}{\partial y^{\mu}}=-\frac{\partial v^{\nu}}{\partial x^{\mu}}\, .
\end{equation}
Es sencillo comprobar que si la acción de $\mathcal{J}_p$ sobre los vectores de la carta $U$ es:
\begin{equation}
\mathcal{J}_p\frac{\partial}{\partial x^{\mu}}=\frac{\partial}{\partial y^{\mu}},~~\mathcal{J}_p\frac{\partial}{\partial y^{\mu}}=-\frac{\partial}{\partial x^{\mu}}\, ,~~~~ \text{entonces}
\end{equation}
\begin{equation}
\mathcal{J}_p\frac{\partial}{\partial u^{\mu}}=\frac{\partial}{\partial v^{\mu}},~~\mathcal{J}_p\frac{\partial}{\partial v^{\mu}}=-\frac{\partial}{\partial u^{\mu}}\, ,
\end{equation}
de manera que la estructura compleja viene dada de forma independiente de la carta por
\begin{equation}
\left(\mathcal{J}_p\right)=\begin{pmatrix}
  0 & \mathbb{I} \\
-\mathbb{I} & 0
 \end{pmatrix},~~\forall p\in \mathcal{M}\, .
\end{equation}
Como en el caso de las estructuras casi complejas, las estructuras complejas permiten descomponer $T_p\mathcal{M}^{\mathbb{C}}$ en sus partes holomorfa y antiholomorfa, con la diferencia de que ahora la descomposición es única, e independiente de la carta. Si escogemos como bases locales los vectores $\left( \frac{\partial}{\partial z^{\mu}},\frac{\partial}{\partial \bar{z}^{\bar{\nu}}}\right)$, la estructura compleja toma su forma estándar
\begin{equation}
\left(\mathcal{J}_p\right)=\begin{pmatrix}
   i\mathbb{I}&0 \\
  0 & -i\mathbb{I}
 \end{pmatrix},~~\forall p\in \mathcal{M}\, .
\end{equation}
Todas las componentes de una $k$-forma $\alpha\in (T_p^*\mathcal{M}\wedge\overset{k}{...}\wedge  T_p^*\mathcal{M})$ en una variedad compleja pueden proyectarse mediante los proyectores $\mathcal{P}^{\pm}$ en sus partes holomorfa y antiholomorfa, de forma que $\alpha$ puede descomponerse en una suma de $k$-formas con $p$ índices holomorfos y $q$ antiholomorfos $\alpha^{(p,q)}$ con $p+q=k$, llamadas $(p,q)$-formas
\begin{equation}
\alpha=\sum_{p+q=k}\alpha^{(p,q)},~~~~\alpha^{(p,q)}=\frac{1}{p!q!}\alpha_{\mu_1 ... \mu_p \bar{\nu}_1... \bar{\nu}_q} dz^{\mu_1}\wedge ... \wedge dz^{\mu_p}\wedge d\bar{z}^{\bar{\nu}_1}...\wedge d\bar{z}^{\bar{\nu}_q}\, .
\end{equation}
Así mismo, en una variedad compleja la derivada exterior $d$ puede descomponerse en los \textit{operadores de Dolbeault} $\partial , \bar{\partial}$
\begin{equation}
d=\partial+\bar{\partial}\, ,
\end{equation}
que actúan según
\begin{equation}
\partial \alpha^{(p,q)}=\alpha^{(p+1,q)}\, ,~~\bar{\partial} \alpha^{(p,q)}=\alpha^{(p,q+1)}\, ,
\end{equation}
donde
\begin{equation}
\alpha^{(p+1,q)}=\frac{1}{p!q!}\partial_{\mu}\alpha^{(p,q)}_{\mu_1 ... \mu_p \bar{\nu}_1... \bar{\nu}_q} dz^{\mu}\wedge dz^{\mu_1}\wedge ... \wedge dz^{\mu_p}\wedge d\bar{z}^{\bar{\nu}_1}...\wedge d\bar{z}^{\bar{\nu}_q}\, .
\end{equation}
\begin{equation}
\alpha^{(p,q+1)}=\frac{1}{p!q!}\partial_{\bar{\nu}}\alpha^{(p,q)}_{\mu_1 ... \mu_p \bar{\nu}_1... \bar{\nu}_q}  dz^{\mu_1}\wedge ... \wedge dz^{\mu_p}\wedge d\bar{z}^{\bar{\nu}}\wedge d\bar{z}^{\bar{\nu}_1}...\wedge d\bar{z}^{\bar{\nu}_q}\, .
\end{equation}
Dichos operadores satisfacen las siguientes propiedades
\begin{equation}
\partial^2=\partial \bar{\partial}+\bar{\partial}\partial=\bar{\partial}^2=0\, .
\end{equation}

\subsection{Variedades hermíticas}

\textit{Definición 6.}\\
Una variedad compleja y riemanniana $(\mathcal{M},\mathcal{G} ,\mathcal{J})$ es una \textit{variedad hermítica} si $\mathcal{G}$ satisface
\begin{equation}
\mathcal{G}_p\left(\mathcal{J}_p u,\mathcal{J}_p v \right)=\mathcal{G}_p\left(u,v \right)~~\forall p\in \mathcal{M},~~u,v\in T_p\mathcal{M}\, ,
\end{equation}
es decir, si $\mathcal{G}$ preserva la estructura compleja $\mathcal{J}$. En tal caso, $\mathcal{G}$ es una \textit{métrica hermítica}.\\

Una variedad compleja siempre admite una métrica hermítica. En efecto, puesto que toda variedad paracompacta $\mathcal{M}$ admite siempre una métrica riemanniana $g$, en todo caso es posible construir otra métrica $\mathcal{G}$
\begin{equation}
\mathcal{G}_p\left(u,v \right)\equiv g_p\left(u,v \right)+g_p\left(\mathcal{J}_pu,\mathcal{J}_pv \right)~~\forall p\in \mathcal{M},~~u,v\in T_p\mathcal{M}\, ,
\end{equation}
que trivialmente preserva la estructura compleja, y por tanto es una métrica hermítica. Las propiedades siguientes se cumplen:
\begin{itemize}
\item $\mathcal{J}_p u$ es ortogonal a $u~~ \forall u\in T_p\mathcal{M}$:
\begin{equation}
\mathcal{G}_p\left(\mathcal{J}_p u,u \right)=\mathcal{G}_p\left(\mathcal{J}_p^2u,\mathcal{J}_p u \right)=-\mathcal{G}_p\left(u,\mathcal{J}_p u \right)=-\mathcal{G}_p\left(\mathcal{J}_pu,u\right)=0\, ,
\end{equation}
\item En la base $\left( \frac{\partial}{\partial z^{\mu}},\frac{\partial}{\partial \bar{z}^{\bar{\nu}}}\right)$ de $T_p\mathcal{M}^{\mathbb{C}}$, las componentes de $\mathcal{G}$ se escriben
    \begin{equation}
\mathcal{G}_{\mu\nu}\equiv \mathcal{G}\left( \frac{\partial}{\partial z^{\mu}},\frac{\partial}{\partial z^{\nu}}\right) =\mathcal{G}\left( \mathcal{J}_p\frac{\partial}{\partial z^{\mu}},\mathcal{J}_p\frac{\partial}{\partial z^{\nu}}\right)=-\mathcal{G}\left( \frac{\partial}{\partial z^{\mu}},\frac{\partial}{\partial z^{\nu}}\right)=0\, ,
\end{equation}
    \begin{equation}
\mathcal{G}_{\bar{\mu}\bar{\nu}}\equiv \mathcal{G}\left( \frac{\partial}{\partial \bar{z}^{\bar{\mu}}},\frac{\partial}{\partial \bar{z}^{\bar{\nu}}}\right) =\mathcal{G}\left( \mathcal{J}_p\frac{\partial}{\partial \bar{z}^{\bar{\mu}}},\mathcal{J}_p\frac{\partial}{\partial \bar{z}^{\bar{\nu}}}\right)=-\mathcal{G}\left( \frac{\partial}{\partial \bar{z}^{\bar{\mu}}},\frac{\partial}{\partial \bar{z}^{\bar{\nu}}}\right)=0\, ,
\end{equation}
\begin{equation}
\mathcal{G}_{\mu\bar{\nu}}\equiv \mathcal{G}\left( \frac{\partial}{\partial z^{\mu}},\frac{\partial}{\partial \bar{z}^{\bar{\nu}}}\right) =\mathcal{G}\left( \mathcal{J}_p\frac{\partial}{\partial z^{\mu}},\mathcal{J}_p\frac{\partial}{\partial \bar{z}^{\bar{\nu}}}\right)=\mathcal{G}\left( \frac{\partial}{\partial \bar{z}^{\bar{\nu}}},\frac{\partial}{\partial z^{\bar{\mu}}}\right)=\mathcal{G}_{\bar{\nu}\mu}\, .
\end{equation}
De esta manera, localmente la métrica puede escribirse
\begin{equation}
\mathcal{G}=\mathcal{G}_{\mu\bar{\nu}} dz^{\mu}\otimes d\bar{z}^{\bar{\nu}}+\mathcal{G}_{\bar{\nu}\mu} d\bar{z}^{\bar{\nu}}\otimes dz^{\mu}\, .
\end{equation}
\end{itemize}

\textit{Definición 7.}\\
Sea $(\mathcal{M},\mathcal{G} ,\mathcal{J})$ una variedad hermítica con métrica hermítica $\mathcal{G}$. Se define la \textit{forma de Kähler} $\omega$ ($\in T_p^*\mathcal{M}\wedge T_p^*\mathcal{M}$) según
\begin{equation}
\omega_p(u,v)\equiv \mathcal{G}_p(\mathcal{J}_p u, v)~~ \forall u,v\in T_p \mathcal{M}\, .
\end{equation}
Sus propiedades son:
\begin{itemize}
\item $\omega$ es antisimétrica
\begin{equation}
\omega_p(u,v)=\mathcal{G}_p(\mathcal{J}_p^2 u, \mathcal{J}_pv)=\mathcal{G}_p(- u, \mathcal{J}_pv)=-\mathcal{G}_p(\mathcal{J}_pv, u)=-\omega(u,v)\, .
\end{equation}
\item $\omega$ preserva la estructura compleja
\begin{equation}
\omega_p(\mathcal{J}_p u,\mathcal{J}_p v)=\mathcal{G}_p(\mathcal{J}_p^2 u, \mathcal{J}_pv)=\mathcal{G}_p(- u, \mathcal{J}_pv)=\omega(u,v)\, .
\end{equation}
\item En coordenadas complejas, sus coordenadas se escriben
\begin{equation}
\omega=i \mathcal{G}_{\mu\bar{\nu}} dz^{\mu}\wedge d\bar{z}^{\bar{\nu}}\, .
\end{equation}
\item La forma producto $\omega \wedge ... \wedge \omega$ (dim$\mathcal{M}$ veces) no se anula en ningún punto de la variedad, de forma que constituye una forma de volumen. Por tanto, toda variedad hermítica es orientable. Así mismo, como toda variedad compleja admite una métrica hermítica a través de la cual se puede definir la forma Kähler, se sigue que toda variedad compleja es orientable.
\end{itemize}

\subsection{Variedades de Kähler}

\textit{Definición 8.}\\
Una variedad hermítica $(\mathcal{M},\mathcal{G} ,\mathcal{J})$ se dice \textit{variedad de Kähler} si su forma de Kähler $\omega$ es cerrada: $d\omega=0$. En este caso, $\mathcal{G}$ se dice una \textit{métrica de Kähler}. De esta manera, una variedad de Kähler es una variedad riemanniana ($\mathcal{G}$), compleja ($\mathcal{J}$) y simpléctica ($\omega$), que incorpora de una manera compatible ($\omega(.,.)=\mathcal{G}(\mathcal{J}.,.)$) las tres estructuras básicas de la geometría diferencial. De la definición de forma Kähler se siguen las siguientes propiedades:
\begin{itemize}
\item Puesto que $d\omega=(\partial + \bar{\partial})\omega=0$, y que la expresión local de $\omega$ no contiene términos de la forma $dz^{\mu}\wedge dz^{\nu}$ ni $d\bar{z}^{\bar{\mu}}\wedge d\bar{z}^{\bar{\nu}}$ (y, como consecuencia, $\partial \omega=\bar{\partial} \omega=0$), $\omega$ define la \textit{clase de cohomología de Dolbeault}
\begin{equation}
[\omega]\in H^{(1,1)}_{\bar{\partial}}(\mathcal{M;\mathbb{C}})\, .
\end{equation}
\item En cada carta $U\subset \mathcal{M}$, la solución de la ecuación $d\omega=0$ viene dada en términos de cierta función $\mathcal{K}\in \mathcal{C}^{\infty}(U,\mathbb{R})$, conocida como \textit{potencial de Kähler}, según
\begin{equation}
\label{kke}
\omega=i\partial_{\mu} \partial_{\bar{\nu}}\mathcal{K} (z,\bar{z})dz^{\mu}\wedge d\bar{z}^{\bar{\nu}}\, .
\end{equation}
\item El potencial de Kähler no está unívocamente definido, es decir, hay varios potenciales Kähler que dan lugar a la misma forma Kähler en cada carta. En efecto, si consideramos la \textit{transformación de Kähler}
\begin{equation}
\label{kahlertransform}
\mathcal{K}^{\prime}=\mathcal{K}+f+f^{\prime}\, ,
\end{equation}
con $f$ cualquier función holomorfa:
\begin{equation}
\omega(\mathcal{K}^{\prime})=i\partial \bar{\partial}(\mathcal{K}+f+f^{\prime})=i\partial \bar{\partial}\mathcal{K}=\omega(\mathcal{K})\, .
\end{equation}
\item Las únicas componentes no nulas de la conexión de Levi-Civita y el tensor de Riemann asociados a la métrica de Kähler vienen dados por
\begin{equation}
\Gamma_{\mu\nu}^{\rho}=\mathcal{G}^{\rho \bar{\rho}} \partial_{\mu}\mathcal{G}_{\bar{\rho}\nu}\, ,~~ \Gamma_{\bar{\mu}\bar{\nu}}^{\bar{\rho}}=\mathcal{G}^{\bar{\rho} \rho} \partial_{\bar{\mu}}\mathcal{G}_{\rho\bar{\nu}}\, ,
\end{equation}
\begin{equation}
R_{\mu\bar{\nu}}=\frac{1}{2}\partial_{\mu}\partial_{\bar{\nu}}\left(\log \det \mathcal{G} \right)\, .
\end{equation}
\item La \textit{forma de Ricci} se define según
\begin{equation}
\mathfrak{R}\equiv i R_{\mu\bar{\nu}} dz^{\mu }\wedge d \bar{z}^{\bar{\nu}}\, .
\end{equation}
Como se comprueba fácilmente, esta es una forma real $\bar{\mathfrak{R}}=\mathfrak{R}$ y cerrada $d\mathfrak{R}=0$. $\mathfrak{R}$ define un elemento no trivial del segundo grupo de cohomología de la variedad $[\mathfrak{R}/2\pi]\in H^2(\mathcal{M};\mathbb{R})$ conocido como \textit{primera clase de Chern} de la variedad: $c_1(\mathcal{M})\equiv[\mathfrak{R}/2\pi]$.
\end{itemize}

\subsection{Variedades de Calabi-Yau}

\textit{Definición 9.}\\
Una variedad de Kähler $(\mathcal{M},\omega ,\mathcal{J})$ compacta de dimensión compleja $n$ es una \textit{variedad de Calabi-Yau} si el grupo de holonomía de la métrica de Kähler $\mathcal{G}$ está contenido en SU($n$). Esta condición implica que la primera clase de Chern de la variedad $c_1(\mathcal{M})$ se anula. La implicación inversa no es cierta en general.

\subsection{Variedades de Kähler-Hodge}

\textit{Definición 10.}\\
Un \textit{fibrado vectorial holomorfo} es un fibrado vectorial complejo sobre una variedad compleja $\mathcal{M}$ tal que el espacio total $E$ es una variedad compleja y la proyección $\pi:E\rightarrow \mathcal{M}$ es holomorfa. En particular, las funciones de transición de un fibrado vectorial holomorfo son holomorfas.

\textit{Definición 11.}\\
Un \textit{fibrado lineal} $\pi:\mathcal{L}\rightarrow \mathcal{M}$ es un fibrado vectorial holomorfo de rango $1$ (es decir, uno en el que cada fibra $\pi^{-1}(p),\,\, p\in\mathcal{M}$ es un espacio vectorial de dimensión $1$).

\textit{Definición 12.}\\
Una variedad de Kähler $(\mathcal{M},\omega ,\mathcal{J})$ es una \textit{variedad de Kähler-Hodge} si y solo si existe un fibrado lineal $\pi:\mathcal{L}\rightarrow \mathcal{M}$ tal que $c_1(\mathcal{L})=[\omega]$, es decir, tal que la clase de cohomología de la forma de Kähler es igual a la primera clase de Chern de $\mathcal{L}$. La primera clase de Chern del espacio total $\mathcal{L}$ se construye a partir de la curvatura $[\mathit{D},\mathit{D}]$ asociada a la derivada covariante con respecto a transformaciones Kähler (\ref{kahlertransform}) definida sobre secciones de $\mathcal{L}$, o más en general, sobre secciones de la familia de fibrados de rango 1 $\left\{\mathcal{L}^{(q,\bar{q})},\, q\in \mathbb{R},\, \bar{q}\in \mathbb{R}\right\}$ según vamos a explicar: en la intersección de dos cartas $U_i$ y $U_j$ de $\mathcal{M}$, una sección $\Psi\in \Gamma(\mathcal{L}^{(q,\bar{q})})$ (que se dice de \textit{peso de Kähler ($q,\bar{q}$)}) se relaciona según
\begin{equation}
\Psi_{(i)}=e^{-(qf_{(i,j)}+\bar{q}\bar{f}_{(i,j)})}\Psi_{(j)}\, ,
\end{equation}
mientras que para el potencial de Kähler se tiene
\begin{equation}
\mathcal{K}_{(i)}=\mathcal{K}_{(j)}+f_{(i,j)}+\bar{f}_{(i,j)}\, .
\end{equation}
Definiendo la \textit{conexión de Kähler} $\mathcal{Q}$
\begin{equation}
\mathcal{Q}\equiv (2i)^{-1}(dz^{\mu}\partial_{\mu}\mathcal{K}-d\bar{z}^{\bar{\nu}}\partial_{\bar{\nu}}\mathcal{K})\, ,
\end{equation}
que en la intersección de dos cartas se relaciona según
\begin{equation}
\mathcal{Q}_{(i)}=\mathcal{Q}_{(j)}-\frac{i}{2}\partial f_{(i,j)}\, ,
\end{equation}
es posible definir una derivada covariante (con respecto a la conexión de Kähler y la conexión de Levi-Civita) sobre las secciones $\Psi\in \Gamma(\mathcal{L}^{(q,\bar{q})})$ mediante
\begin{equation}
\mathfrak{D}_{\mu}\equiv \nabla_{\mu}+iq\mathcal{Q}_{\mu}\, ,\mathfrak{D}_{\bar{\nu}}\equiv \nabla_{\bar{\nu}}-i\bar{q}\mathcal{Q}_{\bar{\nu}}\, ,
\end{equation}
siendo $\nabla$ la derivada covariante asociada a la conexión de Levi-Civita sobre $\mathcal{M}$. Así mismo, es posible definir una derivada covariante con respecto a transformaciones de Kähler solamente según:
\begin{equation}
\mathit{D}_{\mu}\equiv \partial_{\mu}+iq\mathcal{Q}_{\mu}\, ,\mathit{D}_{\bar{\nu}}\equiv \partial_{\bar{\nu}}-i\bar{q}\mathcal{Q}_{\bar{\nu}}\, ,
\end{equation}
(de modo que $\mathfrak{D}\sim\mathit{D}+\Gamma$). Es con respecto a la curvatura asociada a $\mathit{D}$ que se define la primera clase de Chern de $\mathcal{L}$. Así pues, en una variedad Kähler-Hodge, la forma de Kähler $\omega$ es igual a la curvatura $[\mathit{D},\mathit{D}]$ excepto por una forma cerrada, que en este caso se anula.

\subsection{Variedades de Kähler especiales}
La geometría de dKähler juega un papel central en la formulación matemática de las teorías de campos supersimétricas. En el caso que nos ocupa, es decir, el de la supegravedad $\mathcal{N}=2$, $d=4$, todas las magnitudes dinámicas relativas a los campos de los multipletes vectoriales se identifican con diferentes estructuras geométricas en un cierto tipo de variedad de Kähler-Hodge conocida como \textit{variedad de Kähler especial de tipo local}:

\textit{Definición 13.}\\
Una variedad de Kähler-Hodge  $\mathcal{L}\xrightarrow{\pi}\mathcal{M}$ es una \textit{variedad de Kähler especial de tipo local} si existe un fibrado $\mathcal{S}\mathcal{V}=\mathcal{S}\mathcal{M}\otimes\mathcal{L}\xrightarrow{\pi}\mathcal{M}$ tal que para cierta sección holomorfa $\Omega \in\Gamma\left(\mathcal{S}\mathcal{V}\right)$ la forma de Kähler viene dada por $\omega = i\partial\bar{\partial}\log \left(i\Omega_M\bar{\Omega}^M\right)$, donde $\mathcal{S}\mathcal{M}\xrightarrow{\pi}\mathcal{M}$ es un fibrado vectorial plano $(2n_{v}+2)$-dimensional con grupo de estructura $\mathrm{Sp}(2n_{v}+2,\mathbb{R})$.\\

\noindent
$\Omega_M\bar{\Omega}^M = \langle \Omega \mid\bar{\Omega}\rangle $ denota el producto interno de las fibras (simpléctico y hermítico)
$M,N,\cdots = 1\cdots 2n_{v}+2$ son \textit{índices simplécticos} que pueden descomponerse en dos conjuntos de índices $\Lambda,\Sigma\cdots=1,\cdots, n_{v}+1$ tales que, por ejemplo, la sección $\Omega$ se escribe $\Omega^M=(\mathcal{X}^{\Lambda},\mathcal{F}_{\Lambda})^{T}$, de forma que tenemos

\begin{equation}
\label{eq:SGDefFund2}
\Omega \equiv
\left(
\begin{array}{c}
\mathcal{X}^{\Lambda}\\
\mathcal{F}_{\Sigma}\\
\end{array}
\right)
\;\; \rightarrow \;\;
\left\{
\begin{array}{lcl}
\langle \Omega \mid\bar{\Omega}\rangle
& \equiv &
\bar{\mathcal{X}}^{\Lambda}\mathcal{F}_{\Lambda}
-\mathcal{X}^{\Lambda}\bar{\mathcal{F}}_{\Lambda}
= -i\ e^{-\mathcal{K}}\, , \\
& & \\
\partial_{\bar{\nu}}\Omega & = & 0 \, ,\\
& & \\
\langle\partial_{\mu}\Omega\mid\Omega\rangle & = & 0 \; .
     \end{array}
  \right.
\end{equation}

\noindent
Resulta conveniente definir una sección simpléctica covariantemente holomorfa $\mathcal{V}=e^{\frac{\mathcal{K}}{2}}\Omega$\footnote{Esta es una sección de un fibrado vectorial diferente, que no puede ser holomorfo, en tanto que $\mathcal{V}$ se relaciona en la intersección de dos cartas a través de funciones de transición no holomorfas.}, que por tanto cumple

\begin{equation}
\label{eq:SGDefFund}
\mathcal{V} \equiv
\left(
\begin{array}{c}
\mathcal{L}^{\Lambda}\\
\mathcal{M}_{\Sigma}\\
\end{array}
\right) \;\; \rightarrow \;\;
\left\{
\begin{array}{lcl}
\langle \mathcal{V}\mid\bar{\mathcal{V}}\rangle
& \equiv &
\bar{\mathcal{L}}^{\Lambda}\mathcal{M}_{\Lambda}
-\mathcal{L}^{\Lambda}\bar{\mathcal{M}}_{\Lambda}
= -i\, , \\
& & \\
\mathfrak{D}_{\bar{\nu}}\mathcal{V} & = & (\partial_{\bar{\nu}}+
{\textstyle\frac{1}{2}}\partial_{\bar{\nu}}\mathcal{K})\mathcal{V} =0 \, ,\\
& & \\
\langle\mathfrak{D}_{\mu}\mathcal{V}\mid\mathcal{V}\rangle & = & 0 \, .
\end{array}
\right.
\end{equation}

\noindent
Si definimos

\begin{equation}
\label{eq:SGDefU}
\mathcal{U}_{\mu}  \equiv \mathfrak{D}_{\mu}\mathcal{V}
=
\left(
\begin{array}{c}
f^{\Lambda}{}_{\mu}\\
h_{\scriptscriptstyle{\Sigma}\, \mu}
\end{array}
\right)\, ,\,\,\,\,
\bar{\mathcal{U}}_{\bar{\nu}} = \overline{\mathcal{U}_{\nu}} \, ,
\end{equation}

\noindent
se sigue que

\begin{equation}
\label{eq:SGProp1}
\begin{array}{rclrcl}
\mathfrak{D}_{\bar{\nu}}\ \mathcal{U}_{\mu}
& = &
\mathcal{G}_{\mu\bar{\nu}}\ \mathcal{V}\,\hspace{2cm} &
\langle\mathcal{U}_{\mu}\mid\bar{\mathcal{U}}_{\bar{\nu}}\rangle
& = &
i\mathcal{G}_{\mu\bar{\nu}} \, , \\
& & & & & \\
\langle\mathcal{U}_{\mu}\mid\bar{\mathcal{V}}\rangle &  = & 0\, , &
\langle\mathcal{U}_{\mu}\mid\mathcal{V}\rangle & = & 0 \, .\\
\end{array}
\end{equation}
Tomando la derivada covariante de la última identidad, encontramos inmediatamente que
$\langle\mathfrak{D}_{\mu}\mathcal{U}_{\nu}\mid\mathcal{V}\rangle = -\langle\
\mathcal{U}_{\nu}\mid \mathcal{U}_{\mu}\rangle$. Es posible demostrar que el lado derecho de esta ecuación es antisimétrico, mientras que el izquierdo es simétrico, de modo que

\begin{equation}
\label{eq:UU}
\langle\mathfrak{D}_{\mu}\mathcal{U}_{\nu}\mid\mathcal{V}\rangle = \langle
\mathcal{U}_{\nu}\mid \mathcal{U}_{\mu}\rangle = 0\, .
\end{equation}
\noindent
La importancia de esta ecuación viene del hecho de que si agrupamos $\mathcal{E}_{\Lambda} = (\mathcal{V},\mathcal{U}_{\mu})$, podemos ver que $\langle \mathcal{E}_{\Sigma}\mid\bar{\mathcal{E}}_{\Lambda}\rangle$ es una matriz no degenerada. Esto nos permite construir un operador identidad para los índices simplécticos, tal que dada una sección de  $\mathcal{A}\in\Gamma\left( E,\mathcal{M}\right)$ tenemos

\begin{equation}
\label{eq:SGSymplProj}
\mathcal{A} = i\langle\mathcal{A}\mid\bar{\mathcal{V}}\rangle \mathcal{V}
-i\langle\mathcal{A}\mid\mathcal{V}\rangle\ \bar{\mathcal{V}}
+i\langle\mathcal{A}\mid\mathcal{U}_{\mu}\rangle\mathcal{G}^{\mu\bar{\nu}}\
\bar{\mathcal{U}}_{\bar{\nu}}
-i\langle\mathcal{A}\mid\bar{\mathcal{U}}_{\bar{\nu}}\rangle
\mathcal{G}^{\mu\bar{\nu}}\mathcal{U}_{\mu} \, .
\end{equation}
\noindent
Como hemos visto, $\mathfrak{D}_{\mu}\mathcal{U}_{\nu}$ es simétrico en $\mu$ y $\nu$. Por otro lado, su producto con $\bar{\mathcal{V}}$ y $\bar{\mathcal{U}}_{\bar{\nu}}$ se anula como consecuencia de las propiedades anteriores. Definamos ahora el objeto de peso Kähler 2

\begin{equation}
\label{eq:SGDefC}
\mathcal{C}_{\mu\nu\rho} \equiv
\langle \mathfrak{D}_{\mu}\ \mathcal{U}_{\nu}\mid \mathcal{U}_{\rho}\rangle
\;\; \rightarrow\;\;
\mathfrak{D}_{\mu}\ \mathcal{U}_{\nu}  =
i\mathcal{C}_{\mu\nu\rho}\mathcal{G}^{\rho\bar{\epsilon}}\bar{\mathcal{U}}_{\bar{\epsilon}} \, ,
\end{equation}

\noindent
donde la última ecuación es consecuencia de (\ref{eq:SGSymplProj}). Como las $\mathcal{U}$'s son ortogonales, sin embargo, puede verse que $\mathcal{C}$ es completamente simétrico en sus tres índices. Además, uno puede demostrar que
\begin{equation}
\label{eq:SGCProp}
\mathfrak{D}_{\bar{\mu}}\ \mathcal{C}_{\nu\rho\epsilon}  = 0\, ,\hspace{1cm}
\mathfrak{D}_{[\mu}\ \mathcal{C}_{\nu]\rho\epsilon} = 0\, .
\end{equation}
\noindent
La así llamada \textit{matriz periodo} $\mathcal{N}$ se define a través de las relaciones
\begin{equation}
\label{eq:SGDefN}
\mathcal{M}_{\Lambda}  = \mathcal{N}_{\Lambda\Sigma} \mathcal{L}^{\Sigma}\, ,
\hspace{1cm}
h_{\Lambda\, \mu}  = \bar{\mathcal{N}}_{\Lambda\Sigma} f^{\Sigma}{}_{\mu} \, .
\end{equation}

\noindent
La relación $\langle\mathcal{U}_{\mu}\mid\overline{\mathcal{V}}\rangle =0$ implica que $\mathcal{N}$ es simétrica, lo que también hace trivial la identidad $\langle\mathcal{U}_{\mu}\mid\mathcal{U}_{\nu}\rangle =0$.

\noindent
Del resto de propiedades básicas (\ref{eq:SGProp1}) encontramos
\begin{eqnarray}
\mathcal{L}^{\Lambda} \Im{\rm m}\mathcal{N}_{\Lambda\Sigma}
\bar{\mathcal{L}}^{\Sigma} & = & -{\textstyle\frac{1}{2}}\, ,
\label{eq:esta}\\
& & \nonumber \\
\mathcal{L}^{\Lambda} \Im{\rm m}\mathcal{N}_{\Lambda\Sigma} f^{\Sigma}{}_{\mu}
& = &
\mathcal{L}^{\Lambda} \Im{\rm m}\mathcal{N}_{\Lambda\Sigma}
\bar{f}^{\Sigma}{}_{\bar{\nu}}
=0\, ,
\label{eq:esa}\\
& & \nonumber \\
f^{\Lambda}{}_{\mu}\ \Im{\rm m}\mathcal{N}_{\Lambda\Sigma}
\bar{f}^{\Sigma}{}_{\bar{\nu}} & = &   -\textstyle{1\over 2}\mathcal{G}_{\mu\bar{\nu}} \, .
\label{eq:esaotra}
\end{eqnarray}
\noindent
Algunas identidades adicionales que pueden derivarse son
\begin{eqnarray}
\label{eq:SGColl1}
(\partial_{\mu}\mathcal{N}_{\Lambda\Sigma}) \mathcal{L}^{\Sigma}
& = &
-2i\Im{\rm m}(\mathcal{N})_{\Lambda\Sigma}\ f^{\Sigma}{}_{\mu} \, , \\
& & \nonumber \\
\label{eq:SGColl2}
\partial_{\mu}\bar{\mathcal{N}}_{\Lambda\Sigma}\ f^{\Sigma}{}_{\nu}
& = &
-2\mathcal{C}_{\mu\nu\rho}\mathcal{G}^{\rho\bar{\rho}}
\Im{\rm m}\mathcal{N}_{\Lambda\Sigma}
\bar{f}^{\Sigma}{}_{\bar{\rho}} \, ,\\
& & \nonumber \\
\label{eq:SGColl3}
\mathcal{C}_{\mu\nu\rho}
& = &
f^{\Lambda}{}_{\mu}f^{\Sigma}{}_{\nu}
\partial_{\rho}\bar{\mathcal{N}}_{\Lambda\Sigma} \, ,\\
& & \nonumber \\
\label{eq:SGColl4}
\mathcal{L}^{\Sigma}\partial_{\bar{\nu}}\mathcal{N}_{\Lambda\Sigma}
& = &
0 \, , \\
& & \nonumber \\
\label{eq:SGColl5}
\partial_{\bar{\nu}}\bar{\mathcal{N}}_{\Lambda\Sigma}\ f^{\Sigma}{}_{\mu}
& = &
2i\mathcal{G}_{\mu\bar{\nu}}\Im{\rm m}\mathcal{N}_{\Lambda\Sigma}
 \mathcal{L}^{\Sigma} \, .
\end{eqnarray}

\begin{equation}
\label{eq:SGImpId}
U^{\Lambda\Sigma} \equiv  f^{\Lambda}{}_{\mu}\mathcal{G}^{\mu\bar{\nu}}
\bar{f}^{\Sigma}{}_{\bar{\nu}}
=
-\textstyle{1\over 2}\Im{\rm m}(\mathcal{N})^{-1|\Lambda\Sigma}
-\bar{\mathcal{L}}^{\Lambda}\mathcal{L}^{\Sigma} \; ,
\end{equation}

\noindent
de la que se sigue $\bar{U}^{\Lambda\Sigma}=U^{\Sigma\Lambda}$.
\noindent
Podemos definir los proyectores
\begin{eqnarray}
\label{eq:projectorg}
\mathcal{T}_{\Lambda} & \equiv & 2i \mathcal{L}_{\Lambda}
=2i\mathcal{L}^{\Sigma}\Im{\rm m}\,
\mathcal{N}_{\Sigma\Lambda}\, ,\\
& & \nonumber \\
\label{eq:projectorm}
\mathcal{T}^{\mu}{}_{\Lambda} & \equiv & -\bar{f}_{\Lambda}{}^{\mu}=
-\mathcal{G}^{\mu\bar{\nu}}
\bar{f}^{\Sigma}{}_{\bar{\nu}}\Im{\rm m}\,
\mathcal{N}_{\Sigma\Lambda}\, .
\end{eqnarray}
\noindent
Utilizando estas definiciones y las propiedades de arriba puede demostrarse
\begin{equation}
  \label{eq:dN}
  \begin{array}{rcl}
\partial_{\mu}\mathcal{N}_{\Lambda\Sigma} & = &
4\mathcal{T}_{\mu(\Lambda}\mathcal{T}_{\Sigma)}\, ,\\
& & \\
\partial_{\bar{\nu}}\mathcal{N}_{\Lambda\Sigma} & = &
4\bar{\mathcal{C}}_{\bar{\nu}\bar{\rho}\bar{\epsilon}}
\mathcal{T}^{\bar{\nu}}{}_{(\Lambda}\mathcal{T}^{\bar{\rho}}{}_{\Sigma)}\, .\\
  \end{array}
\end{equation}

\noindent
Es fácil ver que la primera de las ecuaciones (\ref{eq:SGDefFund2}) junto con la definición de la matriz periodo $\mathcal{N}$ implican la siguiente expresión para el potencial Kähler
\begin{equation}
\label{eq:kpotential}
e^{-\mathcal{K}}= -2\Im{\rm m}\mathcal{N}_{\Lambda\Sigma}
\mathcal{X}^{\Lambda}  \bar{\mathcal{X}}^{\Sigma}\, .
\end{equation}
\noindent
Si ahora asumimos que $\mathcal{F}_{\Lambda}$ depende de  $z^{\mu}$ a través de las $\mathcal{X}$'s, entonces de la última ecuación podemos derivar
\begin{equation}
\partial_{\mu}\mathcal{X}^{\Lambda}
\left[
2\mathcal{F}_{\Lambda}
-\partial_{\Lambda}\left( \mathcal{X}^{\Sigma}\mathcal{F}_{\Sigma}\right)
\right]
= 0 \; .
\end{equation}
\noindent
Si $\partial_{\mu}\mathcal{X}^{\Lambda}$ es invertible es una matriz $n\times \bar{n}$, entonces debemos concluir que
\begin{equation}
\label{eq:Prepot}
\mathcal{F}_{\Lambda}  = \partial_{\Lambda}\mathcal{F}(\mathcal{X}) \, ,
\end{equation}

\noindent
donde $\mathcal{F}$ es una función homogénea de grado 2, llamada \textit{prepotencial}. Haciendo uso del prepotencial y las definiciones  (\ref{eq:SGDefN}), podemos calcular
\begin{equation}
\label{eq:PrepotN}
\mathcal{N}_{\Lambda\Sigma}
=\bar{\mathcal{F}}_{\Lambda\Sigma}
+2i\frac{
\Im{\rm m}\mathcal{F}_{\Lambda\Lambda^{\prime}}
\mathcal{X}^{\Lambda^{\prime}}
\Im{\rm m}\mathcal{F}_{\Sigma\Sigma^{\prime}}\mathcal{X}^{\Sigma^{\prime}}
}
{
\mathcal{X}^{\Omega}\Im{\rm m}\mathcal{F}_{\Omega\Omega^{\prime}}
\mathcal{X}^{\Omega^{\prime}}
}\, .
\end{equation}
A partir de la forma explícita de $\mathcal{N}$, podemos derivar también una representación explícita para $\mathcal{C}$ aplicando (\ref{eq:SGColl4}). Se encuentra
\begin{equation}
\label{eq:PrepC}
\mathcal{C}_{\mu\nu\rho} =
e^{\mathcal{K}}\partial_{\mu}\mathcal{X}^{\Lambda}
 \partial_{\nu}\mathcal{X}^{\Sigma} \partial_{\rho}\mathcal{X}^{\Omega}
\mathcal{F}_{\Lambda\Sigma\Omega} \; ,
\end{equation}

\noindent
de manera que el prepotencial determina todas las estructuras en geometría especial.

Dada una sección holomorfa $\Omega$, no se sigue que un prepotencial asociado haya de existir. Sin embargo, en \cite{Craps:1997gp} se demostró que siempre es posible aplicar una transformación de Sp$(2n_{v}+2,\mathbb{R})$ de forma que tal prepotencial existe.

\newpage
\section{Supergravedad $\mathcal{N}=2,~d=4$}
\label{N=2}
\subsection{Supersimetría y supergravedad }
La \textit{supersimetría} es una hipotética simetría fundamental de la naturaleza concebida en los años 70 en una serie de trabajos seminales de Golfand, Likhtman, Wess, Zumino, Ramond, Neveu, Schwarz, Gervais, Sakita, Haag, Lopuszanski y Sohnius. En cierto sentido que vamos a explicar, la supersimetría unificaría los dos tipos de campos elementales conocidos, es decir, los bosónicos (interacciones) y los fermiónicos (materia).

En \cite{Coleman:1974bu}, Coleman y Mandula demostraron, bajo una serie de asunciones muy generales relacionadas con el cálculo de amplitudes de dispersión en teoría de campos, que cualquier grupo de Lie que contenga al grupo de Poincaré $P$, cuyos generadores $P_{\mu}$ y $J_{\mu\nu}$ satisfacen las relaciones de conmutación
\begin{align}
\label{poincare}
[P_{\mu},P_{\nu}]&=0\\
[P_{\mu},J_{\nu\rho}]&=(\eta_{\mu\nu}P_{\rho}-\eta_{\mu\rho}P_{\nu})\, ,\\
[J_{\mu\nu},J_{\rho\gamma}]&=-(\eta_{\mu\rho} J_{\nu\gamma} +\eta_{\nu\gamma} J_{\mu\rho}-\eta_{\mu\gamma}\, , J_{\nu\rho}-\eta_{\nu\rho} J_{\mu\gamma})\, ,
\end{align}
\noindent
y un grupo de simetría interno $G$ con generadores $T_s$ tal que
\begin{align}
\label{g}
[T_r,T_s]=f_{rs}^t T_t\, ,
\end{align}
\noindent
ha de ser un producto directo de $P$ y $G$, de modo que
\begin{align}
\label{gp}
[P_{\mu},T_s]=[J_{\mu\nu},T_s]=0\, ,
\end{align}
\noindent
con posibles grupos U$(1)$ adicionales. Así mismo probaron que había de ser semisimple.

Obviamente existen en general grupos de Lie que contienen a $P$ y a los grupos de simetría internos $G$ correspondientes de forma no trivial. Sin embargo, de acuerdo con el teorema de Coleman y Mandula, estos grupos predecirían en todo caso física trivial, es decir, la matriz S sería esencialmente nula para todos los procesos. Una posible forma de evitar este teorema \textit{no-go} consiste en considerar una generalización del concepto de álgebra de Lie. En \cite{Golfand:1971iw}, Golfand y Likhtman demostraron que mediante el uso de una generalización del concepto de álgebra de Lie, las llamadas \textit{álgebras de Lie graduadas}\footnote{Véase el apéndice \ref{graded}.}, sí era posible encontrar un grupo que mezclase el de Poincaré y un grupo interno de simetría de forma \textit{no trivial}.

La idea consiste en promocionar el álgebra de Poincaré (más el grupo de simetría interno $G$) a un \textit{álgebra $\mathbb{Z}_2$-graduada} introduciendo 4$\mathcal{N}$ generadores impares (\textit{supercargas}) $Q^L_{\alpha}$, $\bar{Q}^L_{\dot{\alpha}}$ $L=1,...,\mathcal{N}$; $\alpha,\dot{\alpha}=1,2$ que satisfagan ahora relaciones de anticonmutación y se transformen como espinores de Weyl (en las representaciones $(0,1/2)$ y $(1/2,0)$ del álgebra del grupo de Lorentz respectivamente). Las relaciones de conmutación (\ref{poincare}), (\ref{g}), (\ref{gp}) se mantienen, puesto que el sector bosónico ha de seguir satisfaciendo el teorema de Coleman y Mandula. El resto de relaciones son\footnote{Una introducción accesible a la supersimetría puede encontrarse en \cite{Quevedo:2010ui}}
\begin{align}
\label{susy}
[Q^L_{\alpha},J_{\mu\nu}]&=(\sigma_{\mu\nu})_{\alpha}^{\beta}Q^L_{\beta}\, , \\
[Q^L_{\alpha},P_{\mu}]&= [\bar{Q}^L_{\dot{\alpha}},P_{\mu}]=0\, \\
\{Q^L_{\alpha},\bar{Q}_{\dot{\beta} M} \} & =2(\sigma^{\mu})_{\alpha\dot{\beta}}P_{\mu}\delta^L_M \, , \\
\{Q^L_{\alpha},Q^M_{\beta } \} &=\epsilon_{\alpha\beta}Z^{LM}\, ,
\end{align}

\noindent
$Z^{LM}=-Z^{ML}$ son las llamadas \textit{cargas centrales}, que conmutan con todos los generadores del superálgebra. Al subgrupo $H\subset G$ generado por aquellos elementos que no conmutan con las supercargas, es decir, por los $T_r\in G$ tales que
\begin{equation}
[Q^L_{\alpha},T_r]={S_r}^{L}_M Q^M_{\alpha}\neq 0\, ,
\end{equation}
\noindent
se le denomina \textit{grupo de R-simetría} (donde $S_{rM}^L$ son matrices correspondientes a la representación de $H$ bajo la cual se transforman las supercargas). Si $Z^{LM}=0$, este grupo es $H=$U$(\mathcal{N})$, y si $Z^{LM}\neq0$, $H$ resulta ser un subgrupo de U$(\mathcal{N})$. Haag,  Lopuszanski y Sohnius demostraron, bajo hipótesis análogas a las de Coleman y Mandula \cite{Haag:1974qh}, que la única extensión no trivial del grupo de Poincaré que incluye generadores fermiónicos con relaciones de anticonmutación consistía en el álgebra supersimétrica arriba resumida.

Del estudio de las representaciones irreducibles sin masa del superálgebra de Poincaré se deduce que cada multiplete está formado por $2^{\mathcal{N}}$ estados de helicidades $\lambda_0+\frac{k}{2}$, $(k=0,...,\mathcal{N})$, con multiplicidades $\binom{\mathcal{N}}{k}$ y con el mismo número de estados bosónicos y fermiónicos en cada caso. Asumiendo que no existen partículas sin masa con helicidades $|\lambda \geq 2|$ y que solo hay una partícula con $\lambda=2$, se tiene que el número máximo de supersimetrías viene dado por $\mathcal{N}=8$.

La acción de las transformaciones de supersimetría en el espacio de campos viene dada por $\delta_{\epsilon}=\epsilon_L^{\alpha}Q^L_{\alpha}$, donde $\epsilon_L$ es el parámetro fermiónico de las transformaciones. El conmutador de dos transformaciones supersimétricas se relaciona con una derivada espaciotemporal según
\begin{equation}
[\delta_{\epsilon_1},\delta_{\epsilon_2}]=(\bar{\epsilon}_1\gamma^{\mu}\epsilon_2)\partial_{\mu}+...\, 
\end{equation}
\noindent
Genéricamente, las transformaciones generadas por el espinor $\epsilon$ actúan sobre campos bosónicos ($B$) y fermiónicos ($F$) según
\begin{align}
\delta_{\epsilon}B\sim \bar{\epsilon}F\, ,\\
\delta_{\epsilon}F\sim B\epsilon \, .
\end{align}

Una teoría de campos se dice supersimétrica si es invariante bajo la acción de las correspondientes transformaciones supersimétricas. Si permitimos que los parámetros de las transformaciones dependan de las coordenadas espaciotemporales $\epsilon_L(x)$, todos los campos habrán de estar acoplados al campo gravitatorio, y las teorías invariantes bajo la acción de tales transformaciones locales se dirán \textit{teorías de supergravedad}. En efecto, el gaugeo de una supersimetría global requiere de la introducción de un campo sin masa de espín $\frac{3}{2}$ (el gravitino, o campo de Rarita-Schwinger). La cuantización consistente de un campo de ese tipo impone sobre la teoría la invariancia bajo transformaciones locales de supersimetría de la misma manera que la cuantización de un campo de espín $1$ precisa de la simetría gauge usual y la de un campo de espín $2$ impone invariancia bajo transformaciones generales de coordenadas. La composición de dos transformaciones supersimétricas locales produce ahora una traslación local, es decir, una transformación general de coordenadas infinitesimal, lo que conduce de forma casi inmediata a una teoría de gravedad.

Las teorías de supergravedad pueden verse como casos particulares de la teoría de Cartan-Sciama-Kibble \cite{Ortin:2004ms}, que es una generalización de la relatividad general de Einstein que permite el acoplamiento de fermiones al campo gravitatorio\footnote{La RG no permite en principio el acoplamiento de fermiones a la gravedad, puesto que está formulada de forma que solo el grupo de difeomorfismos Diff$(\mathcal{M})$ del espaciotiempo actúa de forma natural sobre el resto de campos, y este carece de representaciones espinoriales.}. Serían por tanto teorías de gravedad con un cierto contenido en materia más o menos constreñido por la imposición de invariancia bajo las correspondientes transformaciones locales de supersimetría
\begin{align}
\delta_{\epsilon}B&\sim \bar{\epsilon}F\, ,\\
\delta_{\epsilon}F&\sim\partial\epsilon+B\epsilon \, .
\end{align}

En las teorías de campos clásicas, los campos bosónicos se transforman en representaciones tensoriales del grupo de Lorentz SO$(1,3)$, de modo que en RG se corresponden con tensores en el espaciotiempo $\mathcal{M}$. Sin embargo, los campos fermiónicos se transforman en representaciones llamadas espinoriales (fundamentales) del recubridor universal del grupo de Lorentz Spin$(1,3)$, y no se corresponden con ninguna sección de los fibrados tangente o cotangente de $\mathcal{M}$. Para incluir fermiones en el espaciotiempo, este ha de admitir (como variedad) una estructura conocida como \textit{fibrado de espín}\footnote{Para más detalles véase por ejemplo \cite{Shahbazi:2013ksa}}, del cual los fermiones serían secciones. En caso de que la variedad admita tal estructura, puede aplicarse el \textit{formalismo de primer orden} de Cartan-Sciama-Kibble. En él, el campo asociado a la gravedad pasa a ser el \textit{vierbein} (o tétrada) $\bf e$ en lugar de la métrica del espaciotiempo $\bf g$, y ambas están relacionadas por $\mathbf{g}=\eta(\mathbf{e},\mathbf{e})$. La elección de una tétrada $\bf e$ hace manifiesta la invariancia de la teoría bajo transformaciones locales de Spin$(1,3)$, y permite definir la \textit{conexión de espín}, que hace posible la definición de derivadas covariantes con respecto a la acción de dicho grupo. Esta es la maquinaria necesaria para incluir fermiones en la teoría. En este formalismo, la acción de Einstein-Hilbert se escribe en términos de la curvatura $\mathbf{R}(\omega)$ según
\begin{equation}
S_{EH}[\mathbf{e},\omega]=\int  \star \mathbf{R}(\omega)\wedge \mathbf{e} \wedge \mathbf{e}\, .
\end{equation}
Los términos cinéticos para los espinores se construyen ahora utilizando la derivada covariante asociada a la conexión de espín $\omega$, que se considera un campo independiente de la teoría: $\bar{F}\displaystyle{\not} \mathcal{D} F$, con $ \mathcal{D}\sim \partial +\omega$. La ecuación de movimiento de $\omega$ resulta ser una condición algebraica que relaciona el resto de campos entre sí.

Como hemos dicho, la invariancia de las distintas teorías de supergravedad bajo la acción de las correspondientes transformaciones de supersimetría locales constriñe el contenido en campos y la estructura geométrica de los distintos multipletes. En este trabajo estaremos particularmente interesados en la teoría de supergravedad $\mathcal{N}=2$, $d=4$ que, según vamos a explicar, aparece como límite de bajas energías de la teoría de cuerdas tipo-IIA compactificada en una variedad de Calabi-Yau de dimensión compleja $3$.

\subsection{Dualidad eléctrico-magnética y covariancia simpléctica}
Consideremos la acción bosónica\footnote{Los contenidos de esta sección están basados en \cite{Gaillard:1981rj, Andrianopoli:1996cm,Shahbazi:2013ksa}. }
\begin{align}
\label{eq:actionbos}
S= \int d^{4}x \sqrt{|g|}\,
\left\{
R
+\mathcal{G}_{ij}(\phi)\partial_{\mu}\phi^{i}\partial^{\mu}\phi^{j}+2\Im m\mathcal{N}_{\Lambda\Sigma}(\phi)F^{\Lambda}{}_{\mu\nu}F^{\Sigma\, \mu\nu}\right.\\ \notag \left.-2\Re e\mathcal{N}_{\Lambda\Sigma}(\phi)F^{\Lambda}{}_{\mu\nu} \star F^{\Sigma\, \mu\nu}
\right\}\, ,
\end{align}
siendo $\mathcal{G}$ definida positiva y $\mathcal{N}$ una matriz compleja con parte imaginaria definida negativa, y los índices $i=1,...,n_s$ y $\Lambda=0,...,n_v$ contando el número de escalares $\phi^i$ y de 1-formas $A^{\Lambda}$ respectivamente (de manera que hay $n_v+1$ $1$-formas y $n_s $ escalares). Como vamos a ver, esta acción cubre el sector bosónico de todas las teorías de supergravedad $d=4$ no gaugeada.

Las ecuaciones de movimiento correspondientes a la métrica, los escalares y las 1-formas resultan ser
\begin{align}
\label{eoB}
\mathcal{E}_{\mu\nu}&=G_{\mu\nu}+\mathcal{G}_{ij}\left[ \partial_{\mu}\phi^i\partial_{\nu}\phi^j -\frac{1}{2}\partial_{\rho}\phi^i\partial^{\rho}\phi^j\right]+8\Im m \mathcal{N}_{\Lambda\Sigma}{F^{\Lambda+}_{\mu}}^{\rho}F^{\Sigma-}_{\nu\rho}=0\, ,\\
\mathcal{E}_i&=\nabla{_\mu}(\mathcal{G}_{ij}\partial^{\mu}\phi^j)-\frac{1}{2}\partial_i \mathcal{G}_{jk}\partial_{\rho}\phi^j\partial^{\rho}\phi^k+\partial_i[\tilde{F}_{\Lambda}^{\nu\mu*}F^{\Lambda}_{\mu\nu}]=0\, , \\
\mathcal{E}_{\Lambda}^{\mu}&={\nabla_{\nu}}{\star} \tilde{F}_{\Lambda}^{\nu\mu}=0\, ,
\end{align}
donde hemos definido el tensor dual $\tilde{F}_{\Lambda}$ según
\begin{equation}
\label{dual}
\tilde{F}_{\Lambda\mu\nu}\equiv - \frac{1}{4\sqrt{|g|}}\frac{\delta S}{\delta\star F^{\Lambda_{\mu\nu}}}=\Re e \mathcal{N}_{\Lambda\Sigma}F^{\Sigma}_{\mu\nu}+\Im m \mathcal{N}_{\Lambda\Sigma}^* F^{\Sigma}_{\mu\nu}.
\end{equation}
Así mismo, las 1-formas satisfacen las identidades de Bianchi
\begin{equation}
\mathcal{B}^{\Lambda\mu}\equiv \nabla_{\nu}\star F^{\Lambda\nu\mu}=0.
\end{equation}
Si definimos el doblete
\begin{equation}
\mathcal{E}^M_{\mu}\equiv \begin{pmatrix}
   \mathcal{B}^{\Lambda}_{\mu} \\ \mathcal{E}_{\Lambda\mu}, \\
  \end{pmatrix}
\end{equation}
las ecuaciones de Maxwell y las identidades de Bianchi se reescriben simplemente como $\mathcal{E}^M_{\mu}=0$, y por tanto admiten como simetría una rotación arbitraria de GL($2n_v+2,\mathbb{R}$) sobre el índice M. O sea
\begin{equation}
\mathcal{E}^M_{\mu}=0\rightarrow {m^M}_N\mathcal{E}^N_{\mu}=0,\, \,  {m^M}_N\in \text{GL}(2n_v+2,\mathbb{R})\, . \\
\end{equation}
Estas transformaciones actúan sobre los vectores de 2-formas $F^{\Lambda}$ y  $\tilde{F}_{\Lambda}$ según
\begin{equation}
F^M_{\mu}\equiv \begin{pmatrix}
   F^{\Lambda}\\ \tilde{F}_{\Lambda}
  \end{pmatrix}
, F^{\prime M}={m^M}_N F^N.\\
\end{equation}
Sin embargo,  $F^{\Lambda}$ y  $\tilde{F}_{\Lambda}$ no son independientes como se ve en la definición de $\tilde{F}_{\Lambda}$, (\ref{dual}). Por tanto, hemos de imponer que tal definición siga siendo válida para $\tilde{F}^{\prime}_{\Lambda}$, es decir, que
\begin{equation}
\label{dual2}
\tilde{F}^{\prime}_{\Lambda\mu\nu}\equiv - \frac{1}{4\sqrt{|g|}}\frac{\delta S^{\prime}}{\delta\star F^{\prime\Lambda_{\mu\nu}}}\, .
\end{equation}
Para que tal condición pueda satisfacerse, es necesario que la matriz $\mathcal{N}$ se transforme bajo GL($2n_v+2,\mathbb{R}$). Por tanto, ha de imponerse una acción sobre los campos escalares que produzca la matriz periodo $\mathcal{N}^{\prime}$ deseada. Para ello, hemos de considerar una transformación $\xi\in $ Diff($\mathcal{M}_{\text{escalar}}$) sobre la variedad escalar $\mathcal{M}_{\text{escalar}}$, así como la existencia de un homomorfismo de grupos (que de hecho es una representación del grupo Diff($\mathcal{M}_{\text{escalar}}$)) según
\begin{equation}
\mathfrak{i}\, : \text{Diff} (\mathcal{M}_{\text{escalar}})\rightarrow \text{GL}(2n_v+2,\mathbb{R})\, ,
\end{equation}
que a cada difeomorfismo $\xi \in  \text{Diff}(\mathcal{M}_{\text{escalar}})$ le asigna una transformación lineal general $\mathfrak{i}(\xi)\in \text{GL}(2n_v+2)$. Esto nos permite definir la acción simultánea de $\xi$ sobre todos los campos de la teoría
\begin{equation}
\left\{\phi,F^M,\mathcal{N}_{\Sigma\Lambda}(\phi) \right\} \overset{\xi}{\rightarrow} \left\{\xi(\phi),({\mathfrak{i}(\xi))^M}_NF^N,\mathcal{N}^{\prime}_{\Sigma\Lambda}  (\xi(\phi))\right\} \, .
\end{equation}
La condición de consistencia (\ref{dual2}) se traduce en que las transformaciones ${m^M}_N$ han de pertenecer al subgrupo Sp($2n_v+2,\mathbb{R}$), de forma que el homomorfismo $\mathfrak{i}$ se reduce a
\begin{equation}
\mathfrak{i}\, : \text{Diff} (\mathcal{M}_{\text{escalar}})\rightarrow \text{Sp}(2n_v+2,\mathbb{R})\, ,
\end{equation}
y en que la matriz periodo ha de transformarse según:
\begin{equation}
\label{peri}
\mathcal{N}^{\prime}  (\xi(\phi))=(A\mathcal{N}(\phi)+B)(C\mathcal{N}(\phi)+D)^{-1}\, ,
\end{equation}
donde $A$, $B$, $C$ y $D$ son las matrices $(n_v+1)\times (n_v+1)$ dimensionales
\begin{equation}
m\equiv \begin{pmatrix} D & C\\ B & A \end{pmatrix}\in \text{Sp}(2n_v+2,\mathbb{R})\, .
\end{equation}
Así pues, siempre que la matriz periodo satisfaga (\ref{peri}), podemos definir estas transformaciones de \textit{dualidad simpléctica eléctrico-magnética} que actúan como simetrías de las ecuaciones de Maxwell y las identidades de Bianchi. Por otro lado, no todas estas transformaciones son simetrías de la acción (\ref{eq:actionbos}) (ni siquiera lo son del sector correspondiente a las 1-formas). Sin embargo, sí podemos hacer que estas transformaciones sean simetrías de todas las ecuaciones de movimiento de la teoría restringiendo los difeomorfismos Diff($\mathcal{M}_{\text{escalar}}$) asociados a las transformaciones de dualidad a aquellos que sean isometrías de la variedad escalar $\mathcal{G}_{ij}$. Así, el homomorfismo $\mathfrak{i}$ se reduce una vez más
\begin{equation}
\label{homoo}
\mathfrak{i}\, : \text{Isometrías} (\mathcal{M}_{\text{escalar}},\mathcal{G}_{ij})\rightarrow \text{Sp}(2n_v+2,\mathbb{R})\, .
\end{equation}

Resumiendo, las transformaciones de dualidad simpléctica, que son simetrías globales de las ecuaciones de movimiento se corresponden con isometrías de la variedad escalar que actúan sobre los escalares como difeomorfismos, y sobre las 1-formas a través del homomorfismo (\ref{homoo}), con la condición de que la matriz periodo satisfaga (\ref{peri}).

Finalmente, es posible comprobar que las simetrías del lagrangiano son isometrías de la variedad escalar con una inmersión diagonal por bloques en el grupo simpléctico Sp$(2n_v+2,\mathbb{R})$ (o sea, con $B=C=0$).

\subsection{Supergravedad $\mathcal{N}=2$, $d=4$}
Una supergravedad $\mathcal{N}=2$, $d=4$ es una teoría de campos invariante bajo la acción de dos transformaciones locales de supersimetría generadas por sendos espinores (de Weyl o Majorana) independientes entre sí. De acuerdo con esta definición, existen teorías de supergravedad que incluyen términos de órdenes arbitrariamente altos en derivadas en su acción. Sin embargo, nos restringiremos a teorías con lagrangianos de hasta dos derivadas, y consideraremos que ninguna de las simetrías globales (como las generadas por el grupo de R-simetría) ha sido \textit{gaugeada}.

El contenido en campos de cualquier supergravedad \textit{clásica}\footnote{En el sentido de no incluir órdenes más altos en derivadas y tener por tanto un sector bosónico que es un caso particular de la relatividad general.} no gaugeada $\mathcal{N}=2$, $d=4$ es el siguiente
\begin{itemize}
\item Un \textit{multiplete de gravedad}: $(\mathbf{e},\bar{\psi}^I,\psi_I,A^0)$, siendo $\mathbf{e}$ la tétrada, $\psi_I$ un doblete de $SU(2)$ (que es el grupo de R-simetría de la teoría) de gravitinos (1-formas), y $A^0$ el \textit{gravifotón} (1-forma).
\item $n_v$ \textit{multipletes vectoriales}: $(A^i,\bar{\lambda}^i_I,\lambda^{iI},z^i)$, donde $A^i$, $i=1,...,n_v$ son 1-formas, $\lambda^i_I$ son espinores (0-formas) y los $z^i$ son escalares complejos (0-formas). Como vamos a ver, los $z^i$ parametrizan una variedad Kähler especial de dimensión $n_v$.
\item $n_h$ \textit{hipermultipletes}: $(\chi_{\alpha},\chi^{\alpha},q^u)$ donde los $\chi_{\alpha}$, $\alpha=1,...,2n_h$ son espinores (0-formas), y los $q^u$ $u=1,...,4n_h$ son escalares reales (0-formas) que parametrizan una \textit{variedad cuaterniónica} \cite{Besse} $4n_h$-dimensional.
\end{itemize}

Todo lagrangiano supersimétrico, y en particular el de la supergravedad $\mathcal{N}=2$ $d=4$ no gaugeada, es invariante bajo la simetría $\mathbb{Z}_2$ que hace $B\rightarrow B,\,\,F\rightarrow -F$. Como consecuencia, la truncación de todos los campos fermiónicos es siempre consistente\footnote{De manera que cualquier solución de una teoría de supergravedad en la que se han truncado los fermiones será una solución de la teoría completa.}. Estaremos interesados en soluciones puramente bosónicas, de forma que de ahora en adelante truncaremos todos los fermiones. La acción correspondiente al sector bosónico de cualquier supergravedad no gaugeada en cuatro dimensiones (y en particular la de $\mathcal{N}=2$ $d=4$) puede escribirse como

\begin{align}
\label{eq:action}
S= \int d^{4}x \sqrt{|g|}\,
\left\{
R
+h_{uv}(q)\partial_{\mu}q^u \partial^{\mu}q^{v}+\mathcal{G}_{i\bar{j}}(z,\bar{z})\partial_{\mu}z^{i}\partial^{\mu}\bar{z}^{\bar{j}}\right.\\ \notag \left.+2\Im_{\Lambda\Sigma}(z,\bar{z})F^{\Lambda}{}_{\mu\nu}F^{\Sigma\, \mu\nu}-2\Re_{\Lambda\Sigma}(z,\bar{z})F^{\Lambda}{}_{\mu\nu} \star F^{\Sigma\, \mu\nu}
\right\}\, ,
\end{align}
\normalsize
\noindent
donde hemos utilizado un único índice simpléctico $\Lambda=(0,i)$ para todos los campos vectoriales $A^{\Lambda}$, $\Lambda=0,...,n_v$. El primer término corresponde al escalar de Ricci asociado a la tétrada $\textbf{e}$ o, equivalentemente, a la métrica del espaciotiempo $\mathbf{g}$\footnote{Habiendo truncado los fermiones, no es necesario mantener la formulación de primer orden de la teoría.}. El segundo es el término cinético de los hiperescalares, que parametrizan una variedad cuaterniónica con métrica $h_{uv}(q)$. Es fácil demostrar que estos campos pueden fijarse a un valor constante $q_u=q^0_u$ de forma consistente, como asumiremos de ahora en adelante\footnote{De hecho, se cree que no existen soluciones regulares de tipo agujero negro con hiperescalares no constantes. }: sus ecuaciones de movimiento no involucran más campos que los propios hiperescalares y en sus transformaciones de supersimetría no aparece ningún término en el que no haya hiperescalares. El tercer sumando es un \textit{modelo $\sigma$ no lineal} correspondiente al término cinético de los escalares complejos de los multipletes vectoriales, que parametrizan una variedad Kähler especial de dimensión $n_v$ con métrica $\mathcal{G}_{i\bar{j}}(z,\bar{z})$. Los dos sumandos restantes se corresponden con los términos cinético y \textit{CP violating-like} correspondientes a los $n_v+1$ campos vectoriales. $\Im_{\Lambda\Sigma}\equiv \Im m \mathcal{N}_{\Lambda\Sigma}$ (que es definida negativa) y $\Re_{\Lambda\Sigma}\equiv \Re e \mathcal{N}_{\Lambda\Sigma}$ son las partes imaginaria y real de una cierta matriz simpléctica dependiente de los escalares $z^i$, $\mathcal{N}_{\Lambda\Sigma}(z,\bar{z})$. Supersimetría impone que esta matriz juegue precisamente el papel de la matriz periodo definida en (\ref{eq:SGDefN}) sobre la variedad Kähler especial parametrizada por los escalares complejos. Como consecuencia, esta matriz satisface (\ref{peri}), de modo que las ecuaciones de movimiento de la supergravedad $\mathcal{N}=2$, $d=4$ no gaugeada son invariantes bajo transformaciones de dualidad simpléctica que explicamos en el apartado anterior.

Truncados hiperescalares y fermiones, la teoría queda completamente determinada por la elección de una sección simpléctica holomorfa del fibrado $\mathcal{SM}$ con grupo de estructura Sp$(2n_v+2,\mathbb{R})$ definido sobre la variedad escalar o, equivalentemente en caso de que exista, por la función holomorfa y homogénea de segundo grado conocida como prepotencial $\mathcal{F}$ y definida en (\ref{eq:Prepot}). A partir de esta, $\mathcal{G}$ y $\mathcal{N}$ pueden construirse utilizando (\ref{eq:kpotential}), (\ref{kke}) y (\ref{eq:PrepotN}).

\newpage
\section{Teoría de cuerdas tipo-IIA en una variedad de Calabi-Yau}
\label{IIA}
La teoría de cuerdas tipo-IIA compactificada a 4 dimensiones en una variedad de Calabi-Yau de dimension compleja 3 (Calabi-Yau \textit{threefold}) con números de Hodge $(h^{1,1},h^{2,1})$ está descrita, a segundo orden en derivadas, por una teoría de supergravedad $\mathcal{N}=2, d=4$ cuyo prepotencial viene dado en términos de una serie infinita alrededor de $\Im{\rm m}z^i\rightarrow \infty$\footnote{Más precisamente, el prepotencial obtenido de la compactificación de la tipo-IIA en un CY es \textit{simplécticamente equivalente al prepotencial (\ref{eq:IIaprepotential}).}}\cite{Candelas:1990pi,Candelas:1990rm,Candelas:1990qd}
\begin{equation}
\label{eq:IIaprepotential}
\mathcal{F} =  -\frac{1}{3!}\kappa^{0}_{ijk} z^i z^j z^k +\frac{ic}{2}+\frac{i}{(2\pi)^3}\sum_{\{d_{i}\}} n_{\{d_{i}\}} Li_{3}\left(e^{2\pi i d_{i} z^{i}}\right) \ \, ,
\end{equation}

\noindent
donde $z^i,~~ i =1,...,n_v=h^{1,1}$ son los escalares de los multipletes vectoriales,  $c=\frac{\chi\zeta(3)}{ (2\pi)^3}$ es una constante que depende del CY de compactificación\footnote{$\chi$ es la característica de Euler de la variedad que, para CY de dimensión compleja 3, viene dada por $\chi=2(h^{1,1}-h^{2,1})$.},  $\kappa^{0}_{ijk}$ son los números de intersección clásicos, $d_{i}\in\mathbb{Z}^{+}$ es un índice de suma $h^{1,1}$-dimensional, y $Li_{3}(x)$ es la tercera función polilogarítmica, definida en el apéndice \ref{sec:polylog}. Los dos primeros términos en el prepotencial provienen de contribuciones perturbativas en la expansión en $\alpha^{\prime}$ a niveles árbol y 4 \textit{loops} respectivamente
\begin{equation}
\label{eq:IIapert}
\mathcal{F}_{\text{P}} =  -\frac{1}{3!}\kappa^{0}_{ijk} z^i z^j z^k +\frac{ic}{2}\ \, ,
\end{equation}

\noindent
mientras que el tercero codifica las correcciones no perturbativas producidas por instantonces de la \textit{world-sheet}. Estas configuraciones se producen a través de inmersiones no triviales de la \textit{world-sheet} en el CY. Los mapas holomorfos de la \textit{world-sheet} (de genus 0\footnote{Los instantones de genus superior a 0 contribuyen con correcciones de orden mayor que 2 en derivadas.}) de la cuerda en los $h^{1,1}$ 2-ciclos del CY se clasifican mediante los números $d_i$, que cuentan el número de enrollamientos de la \textit{world-sheet} alrededor del $i-$ésimo generador del grupo de homología entero de la variedad $H_2(\text{CY},\mathbb{Z})$. Al número de mapas distintos para cada conjunto de $\{d_i\}$ $\left( \equiv \{   d_1,...,d_{h^{1,1}}\} \right)$ o, en otras palabras, al número de instantones de genus 0 se le denota $n_{\{d_i\}}$\footnote{Para más detalles sobre el origen \textit{cuerdoso} del prepotencial puede consultarse, por ejemplo, \cite{Mohaupt:2000mj}.}
\begin{equation}
\label{eq:IIanonpert}
\mathcal{F}_{\text{NP}} = \frac{i}{(2\pi)^3}\sum_{\{d_{i}\}} n_{\{d_{i}\}} Li_{3}\left(e^{2\pi i d_{i} z^{i}}\right) \ \,.
\end{equation}

\noindent
El prepotencial completo puede reescribirse en coordenadas homogéneas $\mathcal{X}^{\Lambda}$, $\Lambda=(0,i)$ como
\begin{align}
\label{eq:IIaprepotentialX}
F(\mathcal{X}) =  -\frac{1}{3!}\kappa^{0}_{ijk} \frac{\mathcal{X}^i \mathcal{X}^j \mathcal{X}^k}{\mathcal{X}^0} +\frac{ic(\mathcal{X}^0)^2}{2}+\frac{i(\mathcal{X}^0)^2}{(2\pi)^3}\sum_{\{d_{i}\}} n_{\{d_{i}\}} Li_{3}\left(e^{2\pi i d_{i} \frac{\mathcal{X}^{i}}{\mathcal{X}^0}}\right) \ \, ,
\end{align}

\noindent
donde los escalares vienen dados por \footnote{Por tanto, este sistema de coordenadas solo es válido lejos de la región $\mathcal{X}^{0}=0$. }
\begin{equation}
\label{eq:scalarsgeneral}
z^i=\frac{\mathcal{X}^{i}}{\mathcal{X}^0} \ \, .
\end{equation}


\newpage
\section{Agujeros negros en supergravedad y teoría de cuerdas}
\label{BHSugra}
\subsection{Mecánica de agujeros negros}
En el marco de la gravitación newtoniana, la concepción de un objeto masivo lo suficientemente denso como para que su velocidad de escape sea tan elevada que ni la luz pueda escapar de su superficie se debe al geólogo inglés John Michell, que mencionó tal posibilidad por primera vez en una carta escrita a Henry Cavendish en 1783\footnote{Conviene remarcar que esta ``versión newtoniana'' de agujero negro apenas comparte propiedades con la de los agujeros negros de la relatividad general. Estas no van mucho más allá de la intuición de que, en cierto sentido, en ambos casos la luz ``no puede escapar'' como consecuencia del campo gravitatorio producido por el grave en cuestión.}. Durante el siglo XIX, la interacción luz-campo gravitatorio estaba lejos de ser entendida, y la idea permaneció en cuarentena hasta los trabajos de Einstein. Poco después de que Einstein formulara su teoría de la relatividad general, Schwarzschild encontró la primera solución a las ecuaciones de la teoría, que describía el campo gravitatorio producido por una distribución esférica de masa. La solución de Schwarzschild, cuyas propiedades físicas fueron entendiéndose a lo largo de los decenios siguientes, y que están aún lejos de ser comprendidas en el regímen en el que los efectos cuánticos han de ser tenidos en cuenta, fue la primera de una larga serie de soluciones de \textit{tipo agujero negro}.

La definición precisa de \textit{agujero negro} en relatividad general es bastante técnica, no es única en general (pues depende de si el espaciotiempo en cuestión satisface tales o cuales propiedades), y requiere de la introducción de una serie de conceptos previos, que vamos a repasar a continuación\footnote{Los contenidos de esta sección están basados en \cite{Wald:1999vt,Hawking:1973uf,Wald:1984rg}}.

En la teoría de la relatividad general de Einstein, un espaciotiempo \cite{Hawking:1973uf,Wald:1984rg} es una clase de equivalencia de los pares $(\mathcal{M},\mathbf{g})$, donde $\mathcal{M}$ es una variedad diferenciable $d$-dimensional y $\mathbf{g}$ una \textit{métrica pseudo-riemanniana} de signatura (lorentziana) $(+,-,...,-)$, y $(\mathcal{M},\mathbf{g})\sim (\mathcal{M}^{\prime},\mathbf{g}^{\prime})$ si $\exists$ un difeomorfismo $f:\mathcal{M}\rightarrow \mathcal{M}^{\prime}$ tal que $f^*\mathbf{g}^{\prime}=\mathbf{g}$, es decir, si ambas variedades son isométricas.

Dos métricas lorentzianas $\mathbf{g}$ y $\tilde{\mathbf{g}}$ definidas sobre $\mathcal{M}$ se dicen \textit{conformes} si existe una función escalar $\Omega$ sobre $\mathcal{M}$ tal que $\tilde{\mathbf{g}}=\Omega^2 \mathbf{g}$. Una \textit{compactificación conforme} de $(\mathcal{M},\mathbf{g})$ consiste en una elección de una métrica $\tilde{\mathbf{g}}$ tal que $(\mathcal{M},\tilde{\mathbf{g}})$ puede embeberse isométricamente en un dominio compacto $U^{\prime}$ de otra variedad lorentziana $(\mathcal{M}^{\prime},\mathbf{g}^{\prime})$.

Un espaciotiempo se dice \textit{asintóticamente simple} si y solo si admite una compactificación conforme y todas las geodésicas de tipo luz de $\mathcal{M} $ comienzan y terminan en $\partial{\mathcal{M}}$.

Un espaciotiempo se dice \textit{asintóticamente plano} si y solo si contiene un subconjunto abierto $U$ isométrico a otro perteneciente a la frontera de la compactificación conforme de un espaciotiempo asintóticamente simple, y el tensor de Ricci se anula sobre $U$. Conceptualmente, un espaciotiempo asintóticamente plano se corresponde con la versión relativista general (en ausencia de constante cosmológica) de un sistema aislado. En un espaciotiempo de este tipo existe una región alejada de todo grave en la cual la curvatura se vuelve arbitrariamente pequeña y la geometría de Minkowski se recupera asintóticamente.

Para un espaciotiempo asintóticamente plano, la región de \textit{agujero negro} $\mathcal{B}\subset \mathcal{M}$ se define como 
\begin{equation}
\mathcal{B}\equiv \mathcal{M}-I^-(\mathcal{I}^+)\, ,
\end{equation}
donde $\mathcal{I}^+$ denota el infinito futuro nulo (es decir, el conjunto de puntos a los que se aproximan asintóticamente las geodésicas nulas que pueden escapar al infinito espacial) y $I^-$ el pasado cronológico\footnote{Es posible dar definiciones similares de agujeros negros para otros tipos de espaciotiempo en los que hay definida una región asintótica, como por ejemplo para aquellos que son asintóticamente $AdS$.}.

El \textit{horizonte de eventos} $\mathcal{H}$ de un agujero negro se define como la frontera de $\mathcal{B}$. Así pues, $\mathcal{H}$ es la frontera del pasado de $\mathcal{I}^+$.
Un agujero negro consiste por tanto en el conjunto de puntos de $\mathcal{M}$ desde los cuales las geodésicas nulas no pueden escapar a infinito. De esta forma, un observador en $\mathcal{B}$ no puede influir causalmente en nada de lo que ocurra fuera del horizonte: ninguna información enviada desde el interior de $\mathcal{H}$ puede escapar del mismo. Similarmente, toda información enviada al agujero negro desde el exterior se pierde irremediablemente y para siempre en su interior.

 $\mathcal{H}$ es una hipersuperficie nula compuesta por geodésicas nulas inextendibles hacia el futuro sin cáusticas, es decir, la \textit{expansión} $\theta$, que mide cuánto se expanden de media las geodésicas infinitesimalmente próximas a otra en una congruencia, de las geodésicas nulas que forman el horizonte no puede volverse infinitamente negativa. Es importante recalcar que $\mathcal{H}$ no posee significado local alguno (de hecho, la curvatura puede en principio ser arbitrariamente pequeña sobre $\mathcal{H}$): la historia completa del futuro del universo ha de conocerse para poder determinar la localización de un horizonte de eventos. 

Si un espaciotiempo $(\mathcal{M},\mathbf{g})$ asintóticamente plano contiene un agujero negro $\mathcal{B}$, este se dice \textit{estacionario} si existe una familia uniparamétrica de isometrías para $(\mathcal{M},\mathbf{g})$ generadas por un campo de Killing $t^{\mu}$ cuyas órbitas son curvas de género tiempo. Así mismo, un espaciotiempo se dice \textit{estático} si es estacionario y, además, existe una hipersuperficie de género espacio $\Sigma$ que es ortogonal a las órbitas de la isometría.

Una superficie de tipo luz $\mathcal{K}$ cuyos generadores coinciden con las órbitas de un grupo uniparamétrico de isometrías (de manera que existe un campo de Killing $\xi^{\mu}$ normal a $\mathcal{K}$) se denomina \textit{horizonte de Killing}. Un resultado debido a Hawking \cite{Hawking:1973uf} establece que para las soluciones de vacío o acopladas a un campo electromagnético de la ecuación de Einstein, el horizonte de eventos de cualquier agujero negro estacionario ha de ser un horizonte de Killing. Ahora, sea $\mathcal{K}$ un horizonte de Killing (no necesariamente un horizonte de eventos de un agujero negro) con campo de Killing normal $\xi^{\mu}$. Puesto que $\nabla^{\mu} \xi^2$ es también normal a $\mathcal{K}$, ambos vectores han de ser proporcionales en cada punto del horizonte. Por tanto, ha de existir una función $\kappa$, sobre $\mathcal{K}$, conocida como \textit{gravedad superficial} de $\mathcal{K}$, definida por la ecuación
\begin{equation}
\nabla^{\mu}\xi^2=-2\kappa \xi^{\mu}\, .
\end{equation}
Es importante recalcar que esta magnitud está definida sobre el horizonte de un agujero negro solamente cuando este ``está en equilibrio'', o sea, cuando es estacionario, de forma que su horizonte de eventos es un horizonte de Killing. $\kappa$ está relacionada con la fuerza (medida en el infinito espacial) que siente una partícula de masa unidad en el horizonte.
 
Un teorema debido a Bardeen, Carter y Hawking \cite{Bardeen:1973gs} establece que si la ecuación de Einstein se cumple para un tensor de energía momento que satisface la \textit{condición de energía dominante}, entonces $\kappa$ ha de ser constante sobre cualquier horizonte de Killing. Este resultado se conoce como \textbf{ley cero de la mecánica de agujeros negros}, en analogía con la ley cero de la termodinámica, que establece que la temperatura a lo largo de un sistema en equilibrio térmico es constante.

La \textbf{primera ley de la mecánica de agujeros negros} \cite{Bardeen:1973gs} es una identidad que relaciona las variaciones en masa $M$ (en analogía con la energía $E$ para el caso termodinámico), área $A$ del horizonte (en analogía con la entropía $S$ para el caso termodinámico según vamos a explicar), momento angular $J$, y carga eléctrica $Q$ de un agujero negro estacionario cuando es perturbado. A primer orden, estas variaciones siempre satisfacen
\begin{equation}
\delta M=\frac{1}{8\pi}\kappa \delta A+ \Omega \delta J + \Phi \delta Q\, ,
\end{equation}
donde $\Omega$ es la velocidad angular, y $\Phi$ el potencial electrostático.

Si la ecuación de Einstein se satisface para un contenido en materia que satisfaga la \textit{condición de energía tipo luz} ($T_{\mu\nu}k^{\mu}k^{\nu} \geq\,  0 \, \forall \, k$ tal que $k^2=0$) y el agujero negro es \textit{fuertemente asintóticamente predecible}, es decir, si existe una región hiperbólica global (es decir, una que contiene una \textit{superficie de Cauchy}, o sea, una superficie que es intersecada por toda curva inextensible de géneros tiempo y luz una y solo una vez) que contenga a $I^-(\mathcal{I}^+)\cup \mathcal{H}$, puede demostrarse que la expansión $\theta$ satisface $\theta\geq 0$ en todos los puntos de $\mathcal{H}$. Como consecuencia,  la suma de las áreas $A$ de los horizontes de eventos de los agujeros negros contenidos en un espaciotiempo nunca puede disminuir con el tiempo, tal y como descubrió Hawking \cite{Hawking:1971tu}
\begin{equation}
\delta A\geq 0\, .
\end{equation}
Esta es la \textbf{segunda ley de la mecánica de agujeros negros}, en analogía a la segunda ley de la termodinámica, que establece que la entropía de un sistema aislado no puede disminuir.

La analogía matemática entre estas tres leyes y las de la termodinámica se rompe con la forma de Planck-Nernst de la tercera ley, que establece que $S\rightarrow 0$ según $T\rightarrow 0$. En efecto, existen agujeros negros \textit{extremos} (o sea, con $\kappa=0$) con área finita $A$. Sin embargo, hay razones para creer que, a diferencia de las otras tres, esta no ha de verse como una ley fundamental de la termodinámica \cite{Wald:1999vt}. En todo caso, la analogía se mantiene en la formulación según la cual no es posible para un sistema termodinámico (agujero negro) alcanzar el cero abstoluto $T\rightarrow 0$ ($\kappa \rightarrow 0$) por medio de un número finito de procesos físicos.

En la correspondencia que estamos formulando, el papel de la energía $E$ lo jugaría la masa del agujero negro $M$, el de la temperatura una cierta constante multiplicada por la gravedad superficial $T \rightarrow \alpha \kappa$, y el de la entropía otra constante multiplicada por el área del horizonte $S \rightarrow \frac{A}{8\pi \alpha}$. 

A pesar de los sugerente de la analogía establecida, la interpretación literal de las propiedades del agujero negro $M$, $A$, $\kappa$ como magnitudes termodinámicas del mismo carece de sentido en RG. Si bien la identificación $E\rightarrow M$ parece apuntar en la dirección correcta, la temperatura de un agujero negro clásico es cero, puesto que no emite radiación alguna, de manera que en principio carece de sentido asignarle una temperatura proporcional a su gravedad superficial. Como consecuencia, es también inconsistente identificar la variable conjugada correspondiente $S$, con el área del horizonte en cuestión.

Cuando se tienen en cuenta efectos cuánticos, la situación resulta cambiar drásticamente. En efecto, según descubrió Hawking en 1974 \cite{Hawking:1974sw}, como resultado de la creación de pares partícula-antipartícula cerca del horizonte de un agujero negro, este sí emite radiación con un espectro de cuerpo negro perfecto, y a temperatura
\begin{equation}
\label{hawkt}
T=\frac{\kappa}{2\pi}\, .
\end{equation}
De esta manera, la gravedad superficial representa después de todo la temperatura física del agujero negro. Así mismo, cobra ahora sentido preguntarse si la entropía del agujero negro es realmente proporcional al área, con la constante de proporcionalidad fijada a $1/4$ a través de (\ref{hawkt})
\begin{equation}
\label{hawks}
S_{\rm bh}=\frac{A}{4}\, .
\end{equation}
Bastante antes de que Hawking descubriera que la temperatura del agujero negro está realmente relacionada con su gravedad superficial, Bekenstein \cite{Bekenstein:1973ur,Bekenstein:1974ax} había ya propuesto una generalización de la segunda ley de la termodinámica para sistemas en los que se incluya un agujero negro como subsistema. La motivación surgía, además de las analogías formales de las propiedades de los agujeros negros y la termodinámica, del hecho de que la absorción de materia por parte de un agujero negro conllevaría una reducción de la entropía total del universo. Para salvar esta dificultad, Bekenstein propuso que los agujeros negros tienen en realidad una entropía propocional a su área, de modo que la segunda ley de la termodinámica se escribiría ahora $\delta S^{\prime}\geq 0$, donde $S^{\prime}\equiv S+S_{\rm bh}$, siendo $S_{\rm bh}$ la entropía del agujero negro, y $S$ la del resto del universo.

En cierto sentido, las leyes de la mecánica de agujeros negros pueden verse simplemente como casos particulares de las leyes de la termodinámica aplicadas a sistemas que contienen agujeros negros. Ahora bien, al igual que las leyes de la física estadística subyacen a las de la termodinámica, es de esperar que la mecánica de agujeros negros esté regida por la dinámica de ciertos grados de libertad microscópicos. La comprensión de la física involucrada en esta dinámica requiere del uso de una teoría cuántica de la gravedad.  En el contexto de la teoría de cuerdas, el acuerdo entre los cálculos macroscópico y microscópico de la entropía del agujero negro resulta, en efecto, producirse para ciertas familias de agujeros negros extremos y casi extremos \cite{Strominger:1996sh,Horowitz:1996fn,Breckenridge:1996is,Maldacena:1996gb,Breckenridge:1996sn,Behrndt:1997gs} como observaron Strominger y Vafa por primera vez \cite{Strominger:1996sh}.

\subsection{Agujeros negros y supersimetría}
En una teoría de supergravedad\footnote{Una introducción interesante a la relación entre agujeros negros y supersimetría puede encontrarse en \cite{Bellucci:2006zz}}, una configuración de campos se dice supersimétrica o \textit{BPS} (de Bogomol'ny-Prasad-Sommefeld) si preserva alguna simetría, es decir, si \cite{Ortin:2010jm}
\begin{align}
\delta_{\epsilon}B&\sim \bar{\epsilon}F=0\, ,\\
\delta_{\epsilon}F&\sim\partial\epsilon+B\epsilon=0 \, ,
\end{align}
para al menos un espinor $\epsilon$. Para configuraciones puramente bosónicas, la primera ecuación se satisface trivialmente, mientras que a la segunda se la conoce como \textit{ecuación del espinor de Killing} y $\epsilon$ se dice un \textit{espinor de Killing}, si existe una solución para la ecuación correspondiente. Si una configuración dada es invariante bajo el máximo número de espinores de Killing independientes, se dice \textit{maximalmente supersimétrica}.

Por medio del formalismo de primer orden podemos acoplar un campo de (Rarita-Schwinger) espín $3/2$, $\Psi_{\mu}$, al campo gravitatorio, siendo la teoría resultante una supergravedad $\mathcal{N}=1$, $d=4$. Si truncamos ahora el gravitino, las soluciones bosónicas de la teoría original (relatividad general) seguirán siendo soluciones de la teoría supersimétrica. En particular, la solución de Schwarzschild
\begin{equation}
\mathbf{g}=\left(1-\frac{2M}{r} \right) dt \otimes dt-\left(1-\frac{2M}{r} \right)^{-1}dr\otimes dr-r^2 (d\theta \otimes d\theta +\sin^2\theta d\phi\otimes d\phi)\, ,
\end{equation}
que es la única solución estática y esféricamente simétrica de la ecuación de vacío de la relatividad general (teorema de Birkhoff mediante), será también solución de la supergravedad $\mathcal{N}=1$, $d=4$. Sin embargo, esta solución no es supersimétrica, puesto que la ecuación de Killing correspondiente
\begin{align}
\delta_{\epsilon}\Psi_{\mu}|_{\text{Schw.}}=0 \, 
\end{align}
carece de soluciones. Es decir, la solución de Schwarzschild rompe todas las supersimetrías. Por otro lado, el espacio de Minkowski es maximalmente supersimétrico, al preservar cuatro supersimetrías, correspondientes a las componentes de un espinor de Majorana en $d=4$. Obviamente, en la región asintótica $r\rightarrow \infty$, las supersimetrías se recuperan para la solución de Schwarzschild, característica que será común a todas las soluciones asintóticamente planas.

Consideremos ahora el agujero negro de Reissner–Nordström, que es la única solución estática y esféricamente simétrica de la acción de Einstein-Maxwell
\begin{equation}
\mathbf{g}=\left(1-\frac{2M}{r}+\frac{q^2}{r^2} \right) dt \otimes dt-\left(1-\frac{2M}{r}+\frac{q^2}{r^2}  \right)^{-1}dr\otimes dr-r^2 (d\theta \otimes d\theta +\sin^2\theta d\phi\otimes d\phi)\, ,
\end{equation}
donde $q$ es la carga eléctrica de la solución. En este caso, y esta será una propiedad que encontraremos en nuestras soluciones, el agujero negro tiene dos horizontes localizados en $r_{\pm}=M\pm\sqrt{M^2-q^2}$: el horizonte interno $r_-$, que es un \textit{horizonte de Cauchy}, y el horizonte de eventos $r_+$.

De acuerdo con la \textit{conjetura del censor cósmico}, no es posible que una singularidad desnuda (es decir, una que no esté cubierta por un horizonte de eventos) pueda producirse dinámicamente por medio de proceso físico alguno (con un contenido de materia razonable) dentro de los límites de la RG. Así pues, la condición $M^2\geq q^2$ ha de cumplirse en cualquier situación físicamente aceptable, para evitar la aparición de una singularidad desnuda. Esta condición se parece notablemente a la cota de Bogomol'ny-Prasad-Sommefeld para la estabilidad de soluciones de tipo solitón en teorías gauge espontáneamente rotas. Cuando la cota se satura, es decir, cuando $M^2=q^2$, los horizontes coinciden, y el agujero negro se vuelve extremo, puesto que
\begin{equation}
T=\frac{\kappa}{2\pi}=\frac{r_+-r_-}{4\pi r_+^2}=\frac{\sqrt{M^2-q^2}}{2\pi^2 r_+^2}=0\, .
\end{equation}
Desde el punto de vista supersimétrico, lo que ocurre es lo siguiente. La teoría de Einstein-Maxwell en $d=4$ puede embeberse en supergravedad $\mathcal{N}= 2$, $d= 4$ mediante la introducción de dos gravitinos $\Psi_{\mu}^L$, $L=1,2$ (una vez más, la inmersión es consistente, sin necesidad de incluir ningún otro campo adicional). Para valores genéricos de $M$ y $q$, el agujero negro de Reissner–Nordström no preserva ninguna de las $8$ supercargas del álgebra local de supersimetría. Sin embargo, la saturación de la cota BPS $M^2=q^2$ hace que la ecuación de Killing
\begin{equation}
\delta_{\epsilon}\Psi_{\mu}|_{\text{RN extr.}}=0 \, 
\end{equation}
pase a tener cuatro soluciones, haciendo de ella una solución ``\textit{$\frac{1}{2}$-BPS}''. Además, la solución admite una interpretación como solución de tipo solitón de la teoría, pues interpola entre dos vacíos maximalmente supersimétricos, a saber, la solución en el horizonte, que es $AdS_2\times S^2$ (asociada al superálgebra $\mathfrak{psu}(1,1|2)$ \cite{Bellucci:2006zz}), y la solución en el infinito espacial, que es Minkowski. 

Una propiedad interesante de las soluciones extremas es que su entropía solo depende de las cargas cuantizadas, lo que posibilita una comparación de la entropía macroscópica de estos agujeros negros con la calculada microscópicamente en teoría de cuerdas. Además, en el caso de los agujeros negro supersimétricos, toda la información relativa a los valores de los campos en el infinito espacial se pierde en el horizonte, dependiendo estos solo de dichas cargas. En ese caso, el sistema pierde toda memoria de su configuración inicial (en el infinito espacial), y es ``atraído'' por ciertas configuraciones fijas conocidas como \textit{atractores}. En los casos extremos no supersimétricos, el sistema también contiene puntos fijos en el horizonte, pero estos dependen en general también de los valores asintóticos de los escalares. Todas estas propiedades son consecuencia directa del \textit{mecanismo de atractores} \cite{Ferrara:1996dd,Ferrara:1997tw,Tripathy:2005qp,Sen:2005wa,Goldstein:2005hq,Bellucci:2007ds,Ferrara:2007qx,Ferrara:2007tu,Ceresole:2007wx,Ferrara:2008hwa,Bellucci:2008jq} para agujeros negros extremos, que viene a establecer que los escalares de cualquier solución extrema de tipo agujero negro de una teoría de este tipo \textit{fluyen} desde valores arbitrarios en el infinito espacial a otros completamente fijos en el horizonte, y que en los casos supersimétricos están determinados únicamente por las cargas cuantizadas del agujero negro. Tal mecanismo es muy general, y resulta funciona para cualquier teoría contenida en la acción (\ref{eq:actionbos}) y en particular, para las teorías de supergravedad. 
\newpage
La forma más general de cualquier métrica estática y esféricamente simétrica solución de (\ref{eq:actionbos}) viene dada por \cite{Ferrara:1997tw} 

\begin{equation}
\label{eq:generalbhmetric}
\begin{array}{rcl}
\mathbf{g}
& = &
e^{2U(\tau)} dt\otimes dt - e^{-2U(\tau)} \gamma_{\underline{m} \underline{n}}
dx^{\underline{m}}\otimes dx^{\underline{n}}\, ,  \\
& & \\
\gamma_{\underline{m}\underline{n}}
dx^{\underline{m}} \otimes dx^{\underline{n}}
& = & \displaystyle
\frac{r_{0}^{2}}{\sinh^{2} r_{0}\tau}\left[ \frac{r_{0}^{2} }{ \sinh^{2}r_{0}\tau} d\tau\otimes d\tau
+
\displaystyle h_{S^2}\right]\, , \\
& & \\
h_{S^2}& = &d\theta \otimes d\theta +\sin^2\theta d\phi\otimes d\phi\, ,
\end{array}
\end{equation}
donde $\tau$ es la coordenada radial y $r_0$ es el \textit{parámetro de no extremalidad} cuando (\ref{g}) se corresponde con un agujero negro. En tal caso, el exterior del horizonte de eventos está cubierto por $\tau \in (-\infty,0)$, con el horizonte en $\tau\rightarrow -\infty$ y el infinito espacial en $\tau \rightarrow 0^-$. El interior del horizonte de Cauchy está cubierto por $(\tau_S,\infty)$, con el horizonte interno en $\tau\rightarrow \infty$, y la singularidad en un cierto $\tau_S$ positivo y finito \cite{Galli:2011fq}. Bajo la asunción de que el espaciotiempo es estático y esféricamente simétrico, todos los campos de la teoría dependen solo de la coordenada $\tau$. Las ecuaciones de Maxwell pueden integrarse explícitamente, de forma que los campos vectoriales pueden obtenerse como funciones de la coordenada radial y de las cargas eléctricas $q_{\Lambda}$ y magnéticas $p^{\Lambda}$. La resolución de las ecuaciones de movimiento (\ref{eoB}) resulta ahora equivalente a la resolución de las ecuaciones unidimensionales para $U(\tau)$ y $\phi^i(\tau)$ que se derivan de la acción (\ref{FGK}) (con escalares reales en general)\footnote{Véase la sección \ref{HFGK}}.

En el límite extremo $r_0\rightarrow 0$, (\ref{eq:generalbhmetric}) viene dada por
\begin{equation}
\mathbf{g}=e^{2U(\tau)} dt\otimes dt-e^{-2U(\tau)}[\delta_{ab}dx^a\otimes dx^b]\, ,
\end{equation}
donde $x^a$ ($a=1,2,3$) son coordenadas cartesianas. Puede demostrarse que en esa situación
\begin{equation}
\lim_{\tau \rightarrow -\infty} e^{-2U}=\frac{A}{4\pi}\lim_{\tau \rightarrow -\infty} \tau^2\, \, , \lim_{\tau \rightarrow -\infty} \tau \frac{d\phi^i}{d\tau}=0\, ,i=1,...,n_v\, ,
\end{equation}
siendo $A$ el área del horizonte, y $n_v$ el número de escalares. Utilizando la ecuación anterior y asumiendo que los escalares no divergen en el horizonte, es fácil demostrar a partir de la ecuación de movimiento unidimensional correspondiente a estos campos que \cite{Shahbazi:2013ksa}
\begin{equation}
\lim _{\tau \rightarrow -\infty}\phi^i=\phi^i_h\, , \,\, \mathcal{G}^{ij}(\phi_h)\partial_jV_{\rm bh}(\phi_h)=0\, ,
\end{equation}
donde $V_{\rm bh}$ es el \textit{potencial de agujero negro} (Véase sección \ref{HFGK}). Asumiendo que la métrica de la variedad escalar es no degenerada, se sigue inmediatamente que 
\begin{equation}
\label{vbla}
\partial_jV_{\rm bh}(\phi_h)=0\, ,
\end{equation}
es decir, que los posibles valores de los escalares en el horizonte se corresponden con puntos críticos del potencial de agujero negro. Si $V_{\rm bh}$ no tiene direcciones planas (algo que no ocurre en general), de forma que (\ref{vbla}) es un sistema compatible de $n_v$ ecuaciones independientes, el valor de todos los escalares queda fijado en el horizonte en términos de las cargas del agujero negro.

Igualmente, en el caso extremo (supersimétrico o no), es posible demostrar \cite{Shahbazi:2013ksa} que la entropía de la solución viene dada por 
\begin{equation}
S=\pi V_{\rm bh}(\phi_h(\mathcal{Q}))=0\, ,
\end{equation}
de modo que esta viene dada exclusivamente en términos de las cargas cuantizadas de la solución.

\newpage
\section{Formalismo H-FGK}
\label{HFGK}
Asumiendo que todos los campos presentes en la acción bosónica de la supergravedad $\mathcal{N}=2$, $d=4$ acoplada a $n_v$  multipletes vectoriales son estáticos y esféricamente simétricos (de forma que solo dependen de la coordenada $\tau$), (\ref{eq:action}) se reduce a la así llamada acción efectiva FGK \cite{Ferrara:1997tw}, dada por

\begin{equation}
\label{FGK}
I_{\text{FGK}}[U,z^i]=\int d\tau \left\{ (\dot{U})^2+\mathcal{G}_{i\bar{j}}\dot{z}^i\dot{\bar{z}}^{\bar{j}}-e^{2U}V_{\rm bh}(z,\bar{z},\mathcal{Q}) \right\}\, ,
\end{equation}

\noindent
junto con la \textit{ligadura hamiltoniana}, asociada a la independencia explícita de $I_{\text{FGK}}$ con respecto a $\tau$
\begin{equation}
\label{haml}
(\dot{U})^2+\mathcal{G}_{i\bar{j}}\dot{z}^i\dot{\bar{z}}^{\bar{j}}+e^{2U}V_{\rm bh}(z,\bar{z},\mathcal{Q})=r_0^2\, .
\end{equation}
\noindent
En las dos expresiones anteriores, $V_{\rm bh}(z,\bar{z},\mathcal{Q})$ es el llamado \textit{potencial de agujero negro}, que se define como
\begin{eqnarray}
\label{VBH}
V_{\rm bh}(z,\bar{z},\mathcal{Q})&\equiv& \frac{1}{2}\mathcal{M}_{MN}(\mathcal{N})\mathcal{Q}^M\mathcal{Q}^N\, ,
\end{eqnarray}
\noindent
donde $\mathcal{Q}^M$ es el vector simpléctico $(2n_v+2)-$dimensional de cargas eléctricas $q$ y magnéticas $p$
\begin{equation}
\left( \mathcal{Q}^{M} \right)
=
\left(
  \begin{array}{c}
   p^{\Lambda} \\ q_{\Lambda} \\
  \end{array}
\right)\, ,
\end{equation}
\noindent
y  $\mathcal{M}_{MN}(\mathcal{N})$ es una matriz simpléctica y simétrica definida en términos de $I
\equiv \Im \mathfrak{m}(\mathcal{N})$ y $R\equiv \Re \mathfrak{e}(\mathcal{N})$ según
\begin{equation}
\label{M}
\left(\mathcal{M}_{MN}(\mathcal{N}) \right)
\equiv
\left(
\begin{array}{cc}
I+RI^{-1}R  & -RI^{-1} \\
& \\
-I^{-1}R & I^{-1} \\
\end{array}
\right)\, .
\end{equation}
\noindent
Como hemos dicho, la no dependencia explícita de $I_{\text{FGK}}$ en $\tau$ hace que el hamiltoniano asociado a la acción efectiva FGK tome un valor constante. En efecto, la reducción dimensional sobre la métrica (\ref{eq:generalbhmetric}) impone que tal constante sea precisamente el cuadrado del parámetro de no extremalidad $r_0^2$.

A través de la reducción dimensional cuyo resultado es la acción efectiva FGK, reducimos el problema de la obtención de soluciones estáticas y esféricamente simétricas de tipo agujero negro para el conjunto de teorías definidas por (\ref{eq:action}), a la resolución de un sistema mecánico unidimensional de $(2n_v+1)$ variables, dadas por los $n_v$ escalares complejos $z^i(\tau)$, más la función de la métrica $U(\tau)$.

El formalismo H-FGK \cite{Meessen:2011aa,Mohaupt:2011aa,Galli:2012ji,Bueno:2013pja} se basa en un cambio de variables de las $(2n_v+1)$ variables físicas, que definen completamente una solución de tipo agujero negro dado un vector de cargas $(\mathcal{Q}^M)=(p^{\Lambda},q_{\Lambda})^T$, a un nuevo conjunto de $(2n_v+2)$ variables reales $(H^M)=(H^{\Lambda},H_{\Lambda})^T$ con las mismas propiedades de transformación que tal vector bajo la acción del grupo de $U$-dualidad contenido en Sp$(2n_v+2,\mathbb{R})$ y que, como veremos, resultan reducirse a funciones armónicas en el espacio euclídeo $\mathbb{R}^3$ para agujeros negros supersimétricos.

Como el número de variables $H$ no coincide con el de grados de libertad físicos, es conveniente definir una nueva variable compleja $X$ según
\begin{eqnarray}
X&\equiv& \frac{1}{\sqrt{2}} e^{U+i\alpha}\, ,
\end{eqnarray}

\noindent
donde hemos introducido un nuevo grado de libertad $\alpha$, que no habrá de poseer contenido físico alguno.

Según hemos explicado en la sección \ref{N=2}, todos los términos cinéticos y acoplamientos de la acción (\ref{eq:action}) con los hiperescalares truncados pueden derivarse en general de un una función holomorfa y homogénea de grado 2 en las coordenadas complejas $\mathcal{X}^{\Lambda}$ llamada \textit{prepotencial} $\mathcal{F}(\mathcal{X})$. De la homogeneidad de $\mathcal{F}(\mathcal{X})$ se sigue que, definiendo
\begin{equation}
\mathcal{F}_{\Lambda}\equiv \frac{\partial \mathcal{F}}{\partial \mathcal{X}^{\Lambda}} ~~~\text{y} ~~~\mathcal{F}_{\Lambda\Sigma}\equiv \frac{\partial^2 \mathcal{F}}{\partial \mathcal{X}^{\Lambda}\partial \mathcal{X}^{\Sigma}},~~~\text{se tiene}~~~\mathcal{F}_{\Lambda}= \mathcal{F}_{\Lambda\Sigma}\mathcal{X}^{\Sigma}\, .
\end{equation}
\noindent
Puesto que $\mathcal{F}_{\Lambda\Sigma}$ es una función homogénea de grado cero y $X$ tiene el mismo peso K\"ahler que la sección covariantemente holomorfa
\begin{equation}
\label{simplsec}
(\mathcal{V}^M)=\left(
  \begin{array}{c}
   \mathcal{L}^{\Lambda} \\ \mathcal{M}_{\Lambda} \\
  \end{array}\right)=e^{\mathcal{K}/2}\left(
  \begin{array}{c}
   \mathcal{X}^{\Lambda} \\ \mathcal{F}_{\Lambda} \\
  \end{array}\right)\, ,
\end{equation}
\noindent
donde $\mathcal{K}$ es el potencial K\"ahler, tenemos también
\begin{equation}
\label{MX}
\frac{\mathcal{M}^{\Lambda}}{X}=\mathcal{F}_{\Lambda\Sigma}\frac{\mathcal{L}^{\Lambda}}{X}\, .
\end{equation}
\noindent
Definiendo ahora los vectores simplécticos $\mathcal{R}^M$ e $\mathcal{I}^M$ como
\begin{equation}
\label{reim}
\mathcal{R}^{M}\equiv \Re \mathfrak{e} (\mathcal{V}^M/X)\, ,~~~~~ \mathcal{I}^{M}\equiv \Im \mathfrak{m} (\mathcal{V}^M/X)\, ,
\end{equation}
\noindent
es posible reescribir (\ref{MX}) en la forma real
\begin{equation}
\label{stab}
\mathcal{R}^{M}=-\mathcal{M}_{MN}(\mathcal{F})\mathcal{I}^{M}\, ,
\end{equation}
\noindent
donde $\mathcal{M}_{MN}(\mathcal{F})$ viene dada por (\ref{M}), pero en esta ocasión como función de la matriz $\mathcal{F}_{\Lambda\Sigma}$ de segundas derivadas del prepotencial, y no de la matriz periodo $\mathcal{N}_{\Lambda\Sigma}$\footnote{Es posible demostrar que ambas se relacionan mediante $\mathcal{M}_{MN}(\mathcal{F})=-\mathcal{M}_{MN}(\mathcal{N})-2\mathsf{W}^{-1}(H_MH_N+\tilde{H}_M\tilde{H}_N)$, donde $\mathsf{W}(H)$ es el potencial hessiano definido en (\ref{hess})}.
Es también inmediato probar la relación
\begin{equation}
d\mathcal{R}^{M}=-\mathcal{M}_{MN}(\mathcal{F})d\mathcal{I}^{M}\, .
\end{equation}
\noindent
De ella, de su inversa, y de las propiedades de simetría de $\mathcal{M}_{MN}$ podemos derivar las siguientes igualdades
\begin{equation}
\frac{\partial \mathcal{I}^M}{\partial \mathcal{R}_N}=\frac{\partial \mathcal{I}^N}{\partial \mathcal{R}_M}=-\frac{\partial \mathcal{R}^M}{\partial \mathcal{I}_N}=-\frac{\partial \mathcal{R}^N}{\partial \mathcal{I}_M}=-\mathcal{M}^{MN}(\mathcal{F})\, .
\end{equation}

Ahora introducimos dos conjuntos de variables duales $H^M$ y $\tilde{H}_M$, y reemplazamos los campos físicos originales por las $(2n_v+2)$ variables reales $H^M(\tau)$
\begin{eqnarray}
\label{hs}
H^M&\equiv& \mathcal{I}^M(X,z,\bar{X},\bar{z}).
\end{eqnarray}

\noindent
Las variables duales $\tilde{H}^M$ pueden identificarse con $\mathcal{R}^M$, que podemos expresar como funciones de las $H^M$ resolviendo (\ref{stab}). Esto nos permite obtener $\mathcal{V}^M/X$ como función de las $H^M$. Los campos físicos pueden recuperarse entonces como
\begin{eqnarray}
z^i=\frac{\mathcal{V}^i/X}{\mathcal{V}^0/X}=\frac{\tilde{H}^i(H)+iH^i}{\tilde{H}^0(H)+iH^0}\, , ~~~~~~~e^{-2U}=\frac{1}{2|X|^2}=\tilde{H}_M(H)H^M\, .
\end{eqnarray}

\noindent
Finalmente, la fase de $X$, $\alpha$, puede obtenerse resolviendo la ecuación
\begin{equation}
\dot{\alpha}=2|X|^2\dot{H}^MH_M-\left[\frac{1}{2i}\dot{z}^i\partial_i\mathcal{K} +c.c. \right]\, .
\end{equation}

La acción efectiva FGK puede reescribirse en términos de las nuevas variables a partir del \textit{potencial hessiano}, que se define como \cite{Meessen:2011aa,Mohaupt:2011aa,Bates:2003vx}
\begin{equation}
\label{hess}
\mathsf{W}(H)\equiv \tilde{H}_M(H)H^M=e^{-2U}\, ,
\end{equation}

\noindent
que es una función homogénea de segundo grado en las $H^M$. Como consecuencia de esta propiedad, se deduce que
\begin{equation}
\label{hHFGK}
\tilde{H}_M=\frac{1}{2}\frac{\partial \mathsf{W}}{\partial H^M}\equiv \frac{1}{2}\partial_M \mathsf{W}\, ,~~~~~H^M=\frac{1}{2}\frac{\partial \mathsf{W}}{\partial \tilde{H}^M}\, .
\end{equation}

Utilizando ahora las relaciones de arriba, así como algunas otras de geometría de K\"ahler especial, es posible reescribir la acción efectiva (\ref{FGK}) y la condición hamiltoniana (\ref{haml}) completamente en términos de las nuevas variables, obteniéndose
\begin{eqnarray}
\label{actionHFGK}
-I_{\text{H-FGK}}[H]&=&\int d\tau \left\{\frac{1}{2}g_{MN}\dot{H}^M\dot{H}^N-V \right\}\, , \\ \notag \\ \label{eq:Hamiltonianconstraint}
r_0^2&=& \frac{1}{2}g_{MN}\dot{H}^M\dot{H}^N+V\, ,
\end{eqnarray}

\noindent
donde hemos definido la ``métrica'' (no invertible)
\begin{equation}
g_{MN}\equiv \partial_M\partial_N \log \mathsf{W}-2\frac{H_MH_N}{\mathsf{W}}\, ,
\end{equation}

\noindent
y el potencial
\begin{equation}
V(H)\equiv\left\{-\frac{1}{4}g_{MN}+\frac{H_M H_N}{2\mathsf{W}^2} \right\}\mathcal{Q}^M\mathcal{Q}^N\, ,
\end{equation}

\noindent
que resulta estar relacionado con el potencial de agujero negro (\ref{VBH}) mediante
\begin{equation}
V_{\rm bh}=-\mathsf{W}V\, .
\end{equation}

Las ecuaciones de movimiento del formalismo H-FGK resultan ser
\begin{equation}
\label{eomHFGK}
g_{MN}\ddot{H}^N+[PQ,M]\dot{H}^P\dot{H}^Q+\partial_M V=0\, ,
\end{equation}

\noindent
donde
\begin{equation}
[PQ,M]\equiv \partial_{(P}g_{Q)M}-\frac{1}{2}\partial_M g_{PQ}\,
\end{equation}

\noindent
es el símbolo de Christoffel de primera especie.

Si contraemos (\ref{eomHFGK}) con $H^P$ y utilizamos las propiedades de homogeneidad de los distintos términos, así como la ligadura hamiltoniana, encontramos la ecuación
\begin{equation}
\label{Ueq}
\tilde{H}_M\left(\ddot{H}^M-r_0^2H^M\right)+\frac{(\dot{H}^MH_M)^2}{\mathsf{W}}=0\, ,
\end{equation}

\noindent
que se corresponde con la ecuación de movimiento de $U$ en el formalismo FGK original.

En el caso supersimétrico, la ausencia de carga NUT queda codificada en la condición \cite{Bellorin:2006xr}
\begin{equation}
\label{condition}
\dot{H}^MH_M=0\, ,
\end{equation}

\noindent
de acuerdo con la asunción de estaticidad de la métrica. Utilizando tal condición, (\ref{Ueq}) toma la forma
\begin{equation}
\label{Ueq2}
\tilde{H}_M\left(\ddot{H}^M-r_0^2H^M\right)=0\, ,
\end{equation}

\noindent
y puede resolverse en el caso extremo asumiendo que las $H^M$ son lineales en $\tau$, es decir
\begin{equation}
\label{extre}
H^M=A^M-\frac{B^M}{\sqrt{2}}\tau\, ,
\end{equation}

\noindent
donde $A^M$ y $B^M$ son constantes de integración a determinar como funciones de las cargas $\mathcal{Q}^M$ y los valores de los escalares en el infinito espacial $z^i_{\infty}$; y en el caso no extremo asumiendo que son combinaciones lineales de funciones hiperbólicas de $r_0\tau$ \cite{Galli:2011fq}\footnote{Es conveniente enfatizar que estos \textit{Ansatze hiperbólico} (\ref{noHextre}), y \textit{armónico} (\ref{extre}) respectivamente no son completamente generales. Pueden consultarse, p.e. \cite{Galli:2010mg,Galli:2012jh,LopesCardoso:2007ky,Gimon:2009gk,Bossard:2012xsa} donde se presentan ejemplos de soluciones que no los satisfacen.}
\begin{equation}
\label{noHextre}
H^M=A^M\cosh(r_0\tau)+B^M\sinh(r_0\tau)\, .
\end{equation}

En el caso supersimétrico, en general, las $H^M$ vienen efectivamente dadas por funciones armónicas en $\mathbb{R}^3$ de la forma
\begin{equation}
\label{Hsusy}
H^M=A^M-\frac{\mathcal{Q}^M}{\sqrt{2}}\tau\, ,
\end{equation}

\noindent
donde $\mathcal{Q}^M$ es precisamente el vector simpléctico de cargas asociado a la solución.

Finalmente, resulta útil aclarar que la condición (\ref{condition}) puede de hecho imponerse sin pérdida de generalidad, como consecuencia de la invariancia de los campos físicos bajo transformaciones locales de Freudenthal (véase \cite{Galli:2012ji}) sobre las variables del H-FGK formalismo. Se trata, en efecto, de una elección de gauge particular, que puede resultar (y ha resultado, de hecho \cite{Meessen:2012su,Galli:2012pt,Bueno:2013psa,Bueno:2012jc}) conveniente a la hora de construir nuevas soluciones, tanto extremas como no extremas.

\newpage
\section{Agujeros negros cuánticos}
\label{QBH}

\subsection{Límite de gran volumen y truncación.}
\label{limtrun}
La forma general del potencial hessiano $\mathsf{W}(H)$ en el caso que nos ocupa (\ref{eq:IIaprepotential}) es extremadamente complicada. Como consecuencia, cualquier intento de resolver las ecuaciones de movimiento correspondientes, o incluso las ecuaciones algebraicas asociadas que se obtendrían haciendo uso del \textit{Ansatz} hiperbólico para las $H^M$ en el caso general parece abocado al fracaso. Para comenzar a acotar el problema, empecemos por considerar una truncación particular. Como veremos, esta nos permitirá obtener nuevas familias de agujeros negros con interesantes e inesperadas propiedades. La truncación en cuestión viene dada por
\begin{equation}
\label{eq:truncation1}
H^0 = H_0 = H_i = 0,~~p^{0} = p_{0} = q_{i} = 0 \, ,
\end{equation}

\noindent
y haremos uso de ella durante toda la tesis. Una truncación consistente requiere que la ecuación de movimiento del campo truncado se satisfaga idénticamente. El conjunto de ecuaciones  (\ref{eomHFGK}) y (\ref{eq:Hamiltonianconstraint}) (teniendo en cuenta (\ref{eq:truncation1})) es no vacío, habida cuenta de que existe una solución independiente del modelo dada por
\begin{equation}
\label{eq:universalsusy}
H^{i} = a^{i}-\frac{p^{i}}{\sqrt{2}}\tau,~~~~r_0=0\, ,
\end{equation}

\noindent
que corresponde a un agujero negro supersimétrico. Por otro lado, las ecuaciones de movimiento del H-FGK formalismo \textit{no entienden de supersimetría}. Estas consisten en un sistema de ecuaciones diferenciales cuya solución puede escribirse como
\begin{equation}
\label{eq:universalsolution1}
H^{M} = H^M\left(a,b\right)\, ,
\end{equation}

\noindent
donde hemos hecho explícita la dependencia en las $2n_{v}+2$ constantes de integración. Es al introducir la solución (\ref{eq:universalsolution1}) en (\ref{eq:Hamiltonianconstraint}) cuando imponemos, a través de $r_0$, una condición particular sobre la extremalidad del agujero negro. Si $r_0=0$, las constantes de integración quedan fijadas de forma que la solución es extrema. En general, no existe una única solución para la cual $r_0=0$, pero sí es cierto que siempre existe una elección correspondiente a una solución supersimétrica. Por tanto, garantizada la existencia de la solución supersimétrica para nuestra truncación particular, es de esperar la existencia de la correspondiente solución no-extrema (\ref{eq:universalsolution1}) de las ecuaciones de movimiento, de la cual la supersimétrica puede obtenerse a través de una elección particular de las constantes de integración que hacen que (\ref{eq:universalsolution1}) satisfaga (\ref{eq:Hamiltonianconstraint}) para $r_0=0$. Concluimos, por tanto, que
\begin{equation}
\label{eq:consistentproof}
\left\{ H^{P} = 0,\mathcal{Q}^P=0 \right\} ~~\Rightarrow \mathcal{E}_{P}=0\, ,
\end{equation}

\noindent
de modo que la truncación de tantas $H$'s como se desee (junto con las cargas asociadas) es consistente siempre y cuando el potencial hessiano $\mathsf{W}\left(H\right)$ siga estando bien definido.

Asumiendo (\ref{eq:truncation1}), las \textit{ecuaciones de estabilización}, que pueden leerse directamente de (\ref{simplsec}) teniendo en cuenta (\ref{reim}) y (\ref{hs}), toman la forma
\begin{equation}
\label{eq:stabi}
\left( \begin{array}{c} iH^i \\\tilde{H}_i  \end{array} \right)=\frac{e^{\mathcal{K}/2}}{X}\left( \begin{array}{c} \mathcal{X}^i \\ \frac{\partial F(\mathcal{X})}{\partial\mathcal{X}^i}  \end{array} \right)
 \, ,
\left( \begin{array}{c} \tilde{H}^0\\0  \end{array}\right)=\frac{e^{\mathcal{K}/2}}{X}\left( \begin{array}{c} \mathcal{X}^0 \\ \frac{\partial F(\mathcal{X})}{\partial\mathcal{X}^0}  \end{array} \right),
\end{equation}

\noindent
mientras que los campos físicos pueden obtenerse a partir de las variables $H^i$ como
\begin{equation}
\label{eq:phyi}
e^{-2U}=\tilde{H}_iH^i \, ,~~z^i=i\frac{H^i}{\tilde{H}^0}\, ,
\end{equation}

\noindent
una vez que $\tilde{H}^0$ y $\tilde{H}^i$ hayan sido determinadas. Para poder despejar $\tilde{H}^0$ en función de las $H^i$, hemos de resolver la (intricada) ecuación
\begin{equation}
\label{eqR0q}
\frac{\partial{F(H)}}{\partial \tilde{H}^0}=0\, ,
\end{equation}

\noindent
donde $F(H)$ hace referencia ahora al prepotencial expresado en términos de las $H^i$
\begin{align}
\label{eq:IIaprepotentialH}
F(H) =  \frac{i}{3!}\kappa^{0}_{ijk} \frac{H^i H^j H^k}{\tilde{H}^0} +\frac{ic(\tilde{H}^0)^2}{2}+\frac{i(\tilde{H}^0)^2}{(2\pi)^3}\sum_{\{d_{i}\}} n_{\{d_{i}\}} Li_{3}\left(e^{-2\pi  d_{i} \frac{H^{i}}{\tilde{H}^0}}\right)  \, .
\end{align}

\noindent
Una vez despejado $\tilde{H}^0$, expresar $\tilde{H}^i$ en términos de las $H^i$ es labor sencilla. En efecto, por (\ref{eq:stabi}) tenemos simplemente
\begin{equation}
\label{eq:rii}
\tilde{H}_i=-i\frac{\partial{F(H)}}{\partial H^i}\, ,
\end{equation}

\noindent
es decir, solo tendríamos que tomar derivadas del prepotencial.

Expandiendo (\ref{eqR0q}) encontramos
\begin{align}
\label{di}
  -\frac{1}{3!}\kappa^{0}_{ijk} \frac{H^i H^j H^k}{({\tilde{H}^0})^3} +c+\frac{1}{4\pi^3}\sum_{\{d_{i}\}} n_{\{d_{i}\}} \left[Li_{3}\left(e^{-2\pi  d_{i} \frac{H^{i}}{\tilde{H}^0}}\right) +Li_{2}\left(e^{-2\pi  d_{i} \frac{H^{i}}{\tilde{H}^0}}\right)\left[\frac{\pi d_iH^i}{\tilde{H}^0} \right] \right]=0\, .
\end{align}

\noindent
Es evidente que despejar $\tilde{H}^0$ de (\ref{di}) se presenta como una tarea extremadamente complicada. Sin embargo, si consideramos el límite de compactificación de gran volumen ($\Im{\rm m} z^i >> 1$), podemos hacer uso de la siguente propiedad de las funciones polilogarítmicas
\begin{equation}
\label{eq:lilimiti}
\lim_{|w|\rightarrow 0}Li_s(w)=w\, , \forall s\in \mathbb{N}\, ,
\end{equation}

\noindent
habida cuenta de que $w=e^{-2\pi d_{i}\Im m z^{i}}\, , \forall \left\{d_i\right\}\in\left(\mathbb{Z}^{+}\right)^{h^{1,1}}$. (\ref{eq:lilimiti}) nos permite reescribir (\ref{di}) como
\begin{align}
\label{di}\notag
  -\frac{1}{3!}\kappa^{0}_{ijk} \frac{H^i H^j H^k}{({\tilde{H}^0})^3} +c+\frac{1}{4\pi^3}\sum_{\{d_{i}\}} n_{\{d_{i}\}} \left[e^{-2\pi  d_{i} \frac{H^{i}}{\tilde{H}^0}}+e^{-2\pi  d_{i} \frac{H^{i}}{\tilde{H}^0}}\left[\frac{\pi d_iH^i}{\tilde{H}^0} \right] \right]= 0\, , ~~\Im{\rm m}z^{i} >> 1\, .
\end{align}

\noindent
Evidentemente, la corrección dominante en este regimen viene dada por $c$, mientras que el cuarto sumando, que será mucho mayor que el tercero ($\pi d_i H^i/\tilde{H}^0>>1$), nos dará el orden siguiente. Despreciando la corrección más pequeña, dada por el tercer sumando, tenemos
\begin{equation}
\label{diiii}
  -\frac{1}{3!}\kappa^{0}_{ijk} \frac{H^i H^j H^k}{({\tilde{H}^0})^3}+c+\frac{1}{4\pi^3}\sum_{\{d_{i}\}} n_{\{d_{i}\}} e^{-2\pi  d_{i} \frac{H^{i}}{\tilde{H}^0}}\left[\frac{\pi d_iH^i}{\tilde{H}^0} \right] = 0\, .
\end{equation}

\noindent
La suma sobre $\{d_i\}$ en (\ref{diiii}) estará dominada en cada caso por un cierto término correspondiente a un vector particular $\left\{\hat{d}_i\right\}$ (y, como consecuencia, a un $n_{\hat{d}_{i}}\equiv \hat{n}$ particular). Quedándonos únicamente con este, encontramos finalmente la ecuación
\begin{equation}
\label{di4}
  -\frac{1}{3!}\kappa^{0}_{ijk} \frac{H^i H^j H^k}{({\tilde{H}^0})^3}+c+\frac{\hat{n}}{4\pi^3}  e^{-2\pi  \hat{d}_{i} \frac{H^{i}}{\tilde{H}^0}}\left[\frac{\pi \hat{d}_iH^i}{\tilde{H}^0} \right] = 0\, .
\end{equation}

\subsection{Agujeros negros cuánticos}
Comencemos por despreciar todas las correcciones no perturbativas que, como hemos visto, están exponencialmente suprimidas en el límite de gran volumen $\Im{\rm m}z^{i} >> 1$. En ese caso, el prepotencial viene dado por (\ref{eq:IIapert}), y $\tilde{H}^0$ y $\tilde{H}^i$ pueden despejarse fácilmente de (\ref{di4}) y (\ref{eq:rii}) respectivamente quedándonos solo con el término cúbico en las $H$'s y $c$. De esas ecuaciones, y haciendo uso de (\ref{eq:phyi}), se tiene
\begin{equation}
\label{eq:hessian}
e^{-2U}=\mathsf{W}(H)=\alpha\left|\kappa^{0}_{ijk} H^i H^j H^k\right|^{2/3}\, ,
\end{equation}

\noindent
donde $\alpha=\frac{\left( 3! c\right)^{1/3}}{2}$ ha de ser una constante positiva si pretendemos que la métrica sea no singular. Por tanto, encontramos que $c>0$ es una condición necesaria para que nuestras soluciones sean regulares por un lado, y para que nuestra truncación sea consistente por el otro. Como hemos argumentado en la sección [\ref{limtrun}], en la medida en que $c>0$ hace que $\mathsf{W}(H)$ esté bien definido, esta es una condición también suficiente para la consistencia de la truncación. El \textit{potencial de agujero negro} correspondiente viene dado por
\begin{equation}
\label{eq:vbh}
V_{\rm bh} = \frac{\mathsf{W}(H)}{4}\partial_{ij}\log\mathsf{W}(H)\mathcal{Q}^{i}\mathcal{Q}^{j}\, ,
\end{equation}

\noindent
mientras que los campos escalares, imaginarios puros, toman la forma
\begin{equation}
\label{eq:scalars}
z^i= i \left(3! c\right)^{1/3} \frac{H^i}{\left(\kappa^{0}_{ijk} H^i H^j H^k\right)^{1/3}} \, ,
\end{equation}

\noindent
y están sujetos a la siguiente condición, que asegura la regularidad del potencial K\"ahler (gauge $\mathcal{X}^0 = 1$)

\begin{equation}
\label{eq:kahlercondition}
\kappa^{0}_{ijk}\Im{\rm m}z^{i}\Im{\rm m}z^{j}\Im{\rm m}z^{k} > \frac{3c}{2}\, .
\end{equation}

\noindent
Sustituyendo (\ref{eq:scalars}) en (\ref{eq:kahlercondition}), obtenemos
\begin{equation}
\label{eq:kahlerconditionII}
c>\frac{c}{4}\ \, ,
\end{equation}

\noindent
que es una identidad (asumiendo $c>0$) y por tanto no impone condición extra alguna sobre los campos escalares. Esto es debido a que el potencial de K\"ahler es de hecho constante evaluado sobre la solución, viniendo dado por
\begin{equation}
\label{eq:kahlerpotential}
e^{-\mathcal{K}}=6c\ \, ,
\end{equation}

\noindent
que, una vez más, está bien definido si $c>0$. Teniendo en cuenta que el volumen del CY es proporcional a $e^{-\mathcal{K}}$, la ecuación (\ref{eq:kahlerpotential}) implica que tal volumen permanece de hecho constante, de forma que el límite $\Im{\rm m}z^{i}\rightarrow \infty$ no corresponde realmente a un régimen de gran volumen para el CY de compactificación. Esta notable propiedad puede verse como una característica puramente cuántica\footnote{Es importante aclarar que el adjetivo \textit{cuántico} no hace referencia en este contexto a propiedades del espaciotiempo, sino de la \textit{world-sheet} \cite{Mohaupt:2000mj}. En ese sentido, a pesar de que tal denominación está muy extendida en la literatura al referirse a este tipo de correcciones, el apelativo \textit{cuerdoso} podría resultar tal vez más apropiado.} de nuestra solución\footnote{Nótese que para poder despreciar consistentemente las correcciones no perturbativas en (\ref{eq:IIaprepotential}), solo necesitamos considerar el régimen $\Im{\rm m}z^{i}>>1$. De esta forma, el hecho de que el volumen del CY no cambie en tal límite no supone una inconsistencia en la aproximación.}. Por otro lado, siempre es posible alcanzar el límite
$\Im{\rm m}z^{i}\gg1$ tomando $c\gg1$, es decir, escogiendo una variedad de Calabi-Yau con característica de Euler suficientemente grande. En tal caso sí estaríamos considerando que el volumen del CY es realmente grande.

Como hemos argumentado, la truncación no es consistente en el límite clásico $c=0$ (en tanto que, por ejemplo, $\mathsf{W}(H)$ no estaría bien definido, o $e^{-\mathcal{K}}$ no sería definido positivo). De este modo, podemos concluir que nuestras soluciones serán \textit{genuinamente} cuánticas, existiendo solo cuando la correspondiente corrección (codificada en $c$) se ha incorporado a la acción.

Por otro lado, encontramos una restricción topológica sobre el CY que podemos utilizar para compactificar la tipo-IIA, y que ha de satisfacerse para que la teoría admita agujeros negros regulares de la clase definida por nuestra truncación. Tal condición se resume en
\begin{equation}
\label{eq:conditionc}
c>0 ~~ \Rightarrow ~~ h^{11}>h^{21} \, .
\end{equation}

\noindent
(\ref{eq:conditionc}) resulta ser una condición bastante severa sobre el CY, particularmente para $h^{11}$ pequeños. En efecto, el número de variedades de Calabi-Yau de dimensión 3 conocidas \cite{Davies:2011fr,Candelas:1987kf,Batyrev:1994hm} que satisfacen $h^{11}>h^{21}$ resulta ser notablemente reducido, especialmente cuando $h^{11}$ toma valores de orden unidad. Habida cuenta de que nuestro objetivo es la constucción explícita de soluciones de tipo agujero negro, y de que el número de campos escalares coincide con $h^{11}$, estaremos interesados en variedades con $h^{11}$ pequeños \cite{Candelas:2008wb,Braun:2010vc,Davies:2011fr,Bueno:2012jc}, que den lugar a modelos computacionalmente tratables\footnote{La existencia de variedades de Calabi-Yau con $h^{11}>0, h^{21}=0$ es bien conocida en la literatura (\textit{rigid CY}) \cite{Candelas:1985en,Strominger:1985it,Candelas:1993nd}. Sin embargo, para $h^{11}$ pequeños, el número de variedades conocidas se reduce drásticamente.}.

En \cite{Bueno:2012jc} llevamos a cabo la construcción explícita de dos nuevas variedades de Calabi-Yau (que hemos llamado $X^{3,1}$ e $Y^{3,1}$) con $h^{11}=3$, $h^{21}=1$, satisfaciendo tanto la restricción topológica, como la imposición práctica de tener que tratar con una teoría con un número moderadamente pequeño de campos escalares. A continuación llevaremos a cabo la construcción de agujeros negros cuánticos (en el sentido preciso definido líneas atrás) para las teorías obtenidas tras la compactificación de la tipo-IIA en ambas variedades.

\subsection{Agujeros negros cuánticos con \texorpdfstring{$h^{11}=3$}{h11 = 3}}
\label{sec:BH3}
En la dos secciones anteriores hemos presentado una truncación particular de las ecuaciones de movimiento de la supergravedad $\mathcal{N}=2, d=4$ en un \textit{background} estático y esféricamente simétrico, que resulta ser consistente solo para valores positivos del coeficiente perturbativo cuántico $c$ (\ref{eq:conditionc}). En las dos próximas secciones vamos a construir explícitamente nuevas soluciones regulares no extremas (y por tanto, no supersimétricas) de tipo agujero negro de la clase definida en la sección anterior. En primer lugar, consideremos las dos variedades de CY construidas en \cite{Bueno:2012jc}. Para ellas, $h^{11}=3$, $h^{21}=1$, y los números de intersección triples resultan venir dados por
\begin{equation*}
    X^{3,1}\Rightarrow\kappa^0_{111} = 48 ~,~ \kappa^0_{222} =\kappa^0_{333} = 8\, ,
\end{equation*}
\begin{equation*}
    Y^{3,1}\Rightarrow\kappa^0_{122} = 6 ~,~ \kappa^0_{222} = 18 ~,~ \kappa^0_{333} = 8\, .
\end{equation*}

\noindent
Para ellos, el potencial hessiano (\ref{eq:hessian}) viene dado, respectivamente, por
\begin{equation}
\label{Wx}
\mathsf{W}(H)=\alpha\left|48\left(H^1\right)^{3}+8 \left[\left(H^2\right)^{3}+ \left(H^3\right)^{3}\right]\right|^{2/3}\, ,
\end{equation}

\begin{equation}
\label{Wy}
\mathsf{W}(H)=\alpha\left|18 \left(H^2\right)^2\left[{H^1}+{H^2}\right]+8\left(H^3\right)^{3}\right|^{2/3}\, .
\end{equation}

\noindent
La forma de proceder ahora es directa: dado el potencial hessiano en términos de las variables $H$, hemos de introducirlo en las ecuaciones del formalismo H-FGK (\ref{eq:Hamiltonianconstraint}) y (\ref{eomHFGK}), y tratar de resolverlas para tales variables\footnote{Huelga decir que esta tarea puede tornarse realmente complicada.}. Por simplicidad asumimos $H^1=s_2 H^2=s_3 H^3 \equiv H$, ($s_{2,3}=\pm 1$), $p^1=p^2=p^3\equiv p$. Para esta configuración particular encontramos la siguiente solución no extrema para cada conjunto de números de intersección
\begin{equation}
H=a \cosh (r_0 \tau) +\frac{b}{r_0}\sinh(r_0 \tau),~~~~ b=s_b \sqrt{r_0^2 a^2+\frac{p^2}{2}}\, ,
\end{equation}

\noindent
donde, de ahora en adelante, $s_b=\pm 1$. Los escalares, por su parte, resultan ser constantes, y vienen dados por
\begin{equation}
z^1=i(3! c)^{1/3} \lambda^{-1/3}=s_{2,3} z^{2,3}\, ,
\end{equation}

\noindent
donde
\begin{align}
&\lambda = \left[48+8(s_2+s_3)\right]~ ~\text{para}~~X^{3,1},\\ \notag
&\lambda= \left[18+18s_2+8s_3\right]~~\text{para}~~Y^{3,1}\, .
\end{align}

\noindent
La forma explícita de la métrica puede obtenerse de (\ref{eq:generalbhmetric}), (\ref{Wx}) y (\ref{Wy}), y resulta ser
\begin{align}
ds^{2}
 &=
\left[\frac{1}{2}(3!c)^{1/3}\left[a \cosh (r_0 \tau) +\frac{b}{r_0}\sinh(r_0 \tau) \right]^2\right]^{-1} dt^2\\ \notag & -\frac{1}{2}(3!c)^{1/3}\left[a \cosh (r_0 \tau) +\frac{b}{r_0}\sinh(r_0 \tau) \right]^2 \left[{\displaystyle\frac{r_{0}^{4}}{\sinh^{4} r_{0}\tau}}d\tau^2
+
{\displaystyle\frac{r_{0} ^{2}}{\sinh^{2}r_{0}\tau}}d\Omega^{2}_{(2)}\right]\, .  \\
\end{align}

\normalsize
Como los escalares son constantes y no dependen de las cargas, no podemos tomar el límite $\Im{\rm m}z^{i}\rightarrow\infty$, que suprime completamente las correcciones no perturbativas. Aún así, el exponente en (\ref{eq:IIanonpert}) es, en ambos casos, de orden
\begin{equation}
\label{eq:ordernonpert}
  2\pi i d_{i} z^{i} \sim -\frac{1}{3}\sum^{3}_{i=1}d_{i},~~~d_{i}\ge 1 \, ,
\end{equation}

\noindent
de forma que las correcciones no perturbativas son, después de todo, pequeñas. En particular, un orden de magnitud más pequeñas que la parte perturbativa del prepotencial $\mathcal{F}_{Pert} \sim 10\cdot\mathcal{F}_{Non-Pert} $.

La solución cae dentro del cono K\"ahler cuando
\begin{equation}
\label{eq:cond1}
    s_2=s_3=-1, ~~ \text{para}~ X^{3,1}\, ,
\end{equation}
\begin{equation}
\label{eq:cond2}
    s_2=-s_3=1, ~~ \text{para}~ Y^{3,1}\, .
\end{equation}

\noindent
Esto puede verificarse explícitamente comprobando que la métrica K\"ahler es definida positiva
\begin{equation}
\mathcal{G}_{ij^*}=\partial_i\partial_{j^*}\mathcal{K}
\end{equation}
al evaluarla sobre la solución. El resultado es que solamente los conjuntos de $\{s_2,s_3\}$ que dan lugar a métricas K\"ahler definidas positivas (y, como consecuencia, a soluciones que caen dentro del cono de K\"ahler) son las mostradas arriba. Estas condiciones sobre los signos de los campos escalares resultan coincidir exactamente con las obtenidas en \cite{Bueno:2012jc} utilizando técnicas diferentes \cite{Candelas:1990rm}.

Imponiendo que el espaciotiempo sea asintóticamente plano, el valor de la constante $a$ queda fijado como

\begin{equation}
a=-s_b \frac{\Im m z^1}{\sqrt{3c}}\, .
\end{equation}

\noindent
Finalmente, la masa y las entropías de los horizontes interno/externo resultan ser
\begin{equation}
\label{eq:mass48}
M=r_0\sqrt{1+\frac{3cp^2}{2r_0^2(\Im{\rm m} z^1)^2}}\, ,
\end{equation}

\begin{equation}
\label{eq:entropy48}
S_{\pm} = r_0^2\pi\left(\sqrt{1+\frac{3cp^2}{2r_0^2 (\Im{\rm m} z^1)^2}}\pm 1  \right)^{2}\, .
\end{equation}
De este modo, el producto de ambas entropías depende solo de la carga
\begin{equation}
\label{eq:entropy48product}
S_{+} S_{-} = \frac{\pi^2\alpha^2}{4}p^{4}\lambda^{4/3}\, .
\end{equation}

Vale la pena comentar que hemos utilizado el \textit{Ansatz} $H^{i}=a^{i}+b^{i}\tau$ en el caso extremo ($r_0=0$) para llevar a cabo la construcción de otras soluciones (algunas de las cuales bastante complicadas funcionalmente) con escalares constantes. Sin embargo, y presumiblemente como consecuencia de la complejidad de los cálculos, no hemos sido capaces de obenter una solución con escalares no constantes para ninguno de los dos modelos analizados en esta sección (algo que sí logramos en la siguiente aún en el caso no extremo). Esto podría sugerir que la presencia de las correcciones perturbativas tiende a producir un efecto estabilizador sobre tales campos.

\subsection{Modelo \texorpdfstring{$STU$}{STU} cuántico}
\label{sec:STUmodel}

En esta sección consideramos un caso especial, el llamado modelo $STU$. Vamos a construir la primera solución (en cualquier modelo) no extrema con escalares no constantes en presencia de correcciones cuánticas existente en la literatura. Para ello, fijamos $n_v=3,~ \kappa^{0}_{123}=1$. De (\ref{eq:hessian}) obtenemos fácilmente la forma del potencial hessiano\footnote{Es conveniente remarcar que no hemos sido capaces de construir un CY explícito con $\kappa^{0}_{123}=1$ y $h^{21}<3$.}
\begin{equation}
\label{eq:truncationSTU}
e^{-2U}=\mathsf{W}(H)=\alpha\left|H^1 H^2 H^3\right|^{2/3}\ \ ,
\end{equation}

\noindent
con $\alpha=3 c^{1/3}$. Los campos escalares vienen dados por
\begin{equation}
\label{eq:scalarsSTU}
z^i= i c^{1/3} \frac{H^i}{\left(H^1 H^2 H^3\right)^{1/3}} \ \ .
\end{equation}

\noindent
De nuevo, la dependencia en $\tau$ de las variables $H$ puede obenterse resolviendo las ecuaciones (\ref{eq:Eqsofmotion}) y (\ref{eq:Hamiltonianconstraint}). La solución resulta ser\footnote{La forma explícita de la métrica puede obtenerse sin más que sustituir (\ref{eq:solutionHSTU}) en (\ref{eq:truncationSTU}) y (\ref{eq:generalbhmetric}).}
\begin{equation}
\label{eq:solutionHSTU}
H^i = a^i \cosh \left(r_0\tau\right) + \frac{b^i}{r_0} \sinh \left(r_0\tau\right), ~~~~ b^i=s^i_b\sqrt{r^2_0 (a^i)^2+\frac{(p^i)^2}{2}}\ \ .
\end{equation}

\noindent
Las tres constantes $a^i$ pueden fijarse relacionándolas con los valores de los escalares en el infinito, e imponiendo que la solución sea asintóticamente plana. Tenemos, así pues, cuatro condiciones para tres parámetros, de forma que podría esperarse una cierta condición sobre los $\Im{\rm m}z^i_{\infty}$, dejando $c$ indeterminada. Sin embargo, el cálculo explícito muestra que la cuarta relación es compatible con las demás, de manera que tal condición extra no es necesaria. Las $a^i$ están dadas por
\begin{equation}
\label{eq:askSTU}
a^i=-s^i_b\frac{\Im{\rm m}z_{\infty}^i}{\sqrt{3 c}}\ \ .
\end{equation}

Por su parte, la masa y las entropías son
\begin{equation}
\label{eq:massSTU}
M=\frac{r_0}{3}\sum_i \sqrt{1+\frac{3c(p^i)^2}{2r_0^2(\Im{\rm m} z_{\infty}^i)^2}}\ \ ,
\end{equation}
\begin{equation}
\label{eq:entropySTU}
S_{\pm} = r_0^2\pi\prod_i \left(\sqrt{1+\frac{3c(p^i)^2}{2r_0^2 (\Im{\rm m} z_{\infty}^i)^2}}\pm 1 \right)^{2/3}\ \ ,
\end{equation}

\noindent
de forma que, nuevamente, el producto de las entropías de los dos horizontes del agujero negro solo depende de las cargas
\begin{equation}
\label{eq:entropySTUproduct}
S_{+} S_{-} = \frac{\pi^2\alpha^2}{4}\prod_i\left(p^i\right)^{4/3}\ \ .
\end{equation}

En el límite extremo ($r_0\rightarrow 0$) recuperamos tanto la solución supersimétrica como la extrema no supersimétrica, dependiendo del signo escogido para las cargas.

\newpage
\section{Agujeros negros no perturbativos y la conjetura de ``no-pelo''}
\label{NPBH}
Las soluciones construidas en la sección anterior \cite{Bueno:2012jc} fueron, junto con las presentadas poco después en \cite{Galli:2012pt} , las primeras de tipo agujero negro no extremo (con escalares constantes y no constantes) de (\ref{eq:IIaprepotential}) en presencia de la corrección cuántica perturbativa codificada en la característica de Euler de la variedad de CY de compactificación, e ignorando las correcciones no perturbativas (mucho más pequeñas que esta cuando $\Im{\rm m}z^{i} >> 1$). Como hemos visto, nuestras soluciones presentaban la particularidad de ser \textit{genuinamente} cuánticas, en tanto que no solo su límite clásico $c\rightarrow 0$ no estaba definido, sino que la propia truncación de la teoría se tornaba inconsistente en tal límite. Se plantea ahora la pregunta de si las contribuciones no perturbativas despreciadas para la construcción de tal familia de soluciones podrían de hecho curar, o al menos mejorar, este comportamiento. Por otro lado, explorar la posible existencia y propiedades de soluciones de tipo agujero negro en presencia de estas correcciones resulta una tarea interesante \textit{per se}. Para afrontar estas cuestiones, restrinjámonos ahora al conjunto de variedades de CY con característica de Euler nula ($c=0$) (las así llamadas variedades de CY \textit{auto-especulares}), de forma que la contribución subdominante en el prepotencial en el régimen de gran volumen no estará dada por el término perturbativo (que ya no estará presente), sino que tendrá un origen no perturbativo.

En la sección \ref{limtrun} mostramos como en el régimen en el que $\Im{\rm m}z^{i} >> 1$, la ecuación para $\tilde{H}^0$, a partir de la cual podíamos obtener las $\tilde{H}^i$, y consecuentemente la forma del potencial hessiano, se reducía, cuando $c=0$, a
\begin{equation}
\label{ecr0}
  -\frac{1}{3!}\kappa^{0}_{ijk} \frac{H^i H^j H^k}{({\tilde{H}^0})^3}+\frac{\hat{n}}{4\pi^3}  e^{-2\pi  \hat{d}_{l} \frac{H^{l}}{\tilde{H}^0}}\left[\frac{\pi \hat{d}_nH^n}{\tilde{H}^0} \right] = 0\, .
\end{equation}

\noindent
De forma más o menos sorprendente, esta ecuación puede resolverse para $\tilde{H}^0$, y la solución resulta ser\footnote{De ahora en adelante, continuaremos utilizando $\mathsf{W}$ para el potencial hessiano, y $W$ para la función de Lambert. Esperamos que esto no dé lugar a confusión.}
\begin{equation}
\label{roW}
\tilde{H}^0=\frac{\pi \hat{d}_lH^l}{ W_a\left(s_a \sqrt{\frac{3\hat{n}(\hat{d}_nH^n)^3}{2\kappa^0_{ijk}H^i H^j H^k}}\right)} \, ,
\end{equation}

\noindent
donde $W_a(x), ~(a=0,-1)$ resulta ser (cualquiera de las dos ramas reales de) la \textit{función $W$ de Lambert}\footnote{Véase el apéndice \ref{sec:lambert} para más detalles.} (también conocida como \textit{logaritmo producto}), y $s_a=\pm1$. Utilizando ahora (\ref{roW}) y (\ref{eq:rii}) podemos despejar $\tilde{H}^i$. El resultado es
\begin{equation}
\tilde{H}_i=\frac{1}{2}\kappa^{0}_{ijk} \frac{H^j H^k}{\pi \hat{d}_lH^l}W_a\left(s_a \sqrt{\frac{3\hat{n}(\hat{d}_mH^m)^3}{2\kappa^0_{pqr}H^p H^q H^r}}\right)\, .
\end{equation}

Los campos físicos pueden escribirse ahora como función de las variables $H$ como
\begin{equation}\label{hessiano}
e^{-2U}=\mathsf{W}(H)=\frac{\kappa^{0}_{ijk} H^i H^j H^k}{2\pi \hat{d}_m H^m} W_a\left(s_a\sqrt{\frac{3\hat{n}(\hat{d}_lH^l)^3}{2\kappa^0_{pqr}H^p H^q H^r}}\right)\, ,
\end{equation}
\begin{equation}\label{scalar}
z^i=i\frac{H^i}{\pi \hat{d}_mH^m}W_a\left(s_a \sqrt{\frac{3\hat{n}(\hat{d}_lH^l)^3}{2\kappa^0_{pqr}H^p H^q H^r}}\right)\, .
\end{equation}

\noindent
Para poder obtener una solución regular, es necesario que el factor de la métrica $e^{-2U}$ sea positivo definido. Habida cuenta de que, según se explica en el apéndice \ref{sec:lambert}, ${\rm sign} \left[W_a(x)\right]={\rm sign}\left[x\right],~a=0,-1,~x\in D^a_{\mathbb{R}}$, hemos de imponer que
\begin{equation}
s_0\equiv sign\left[\kappa^{0}_{ijk} \frac{H^i H^j H^k}{ \hat{d}_m H^m} \right]\, ,
\end{equation}
\begin{equation}
s_{-1}\equiv -1\, .
\end{equation}

\noindent
Por otro lado, como $W_{0}(x)=0$ para $x=0$ y $W_{-1}(x)$ es una función real solamente cuando $x\in\left[-\frac{1}{e},0\right)$, hemos de imponer sobre el argumento $x$ de $W_{a}$ la condición de que caiga enteramente o bien en un intervalo contenido en $\left[-\frac{1}{e},0\right)$, o en uno contenido en $\left(0,+\infty\right)$ para todo $\tau\in\left(-\infty,0\right)$, puesto que $e^{-2U}$ no puede anularse en ninguna solución regular de tipo agujero negro para ningún $\tau\in\left(-\infty,0\right)$. Esta condición ha de imponerse, en principio, caso por caso, puesto que depende de la forma específica de las $H$ como función de $\tau$. Por otro lado, si $x\in \left[-\frac{1}{e},0\right)~~\forall~~\tau\in\left(-\infty,0\right)$, podemos en principio\footnote{Como veremos en la sección \ref{sec:susysolution}, la posibilidad $s_0=s_{-1}=-1$ no será consistente con la aproximación de gran volumen que estamos considerando.} elegir entre $W_{0}$ y $W_{-1}$ para construir nuestra solución, mientras que si $x\in \left(0,+\infty \right)~~\forall~~\tau\in\left(-\infty,0\right)$, solo $W_0$ tomará valores reales.

Evidentemente, una vez más, para construir soluciones hemos de resolver las ecuaciones del formalismo H-FGK (\ref{eomHFGK}) y (\ref{eq:Hamiltonianconstraint}) utilizando el potencial hessiano dado por (\ref{hessiano}). En este caso, este potencial resulta ser bastante complicado funcionalmente, por lo que la manipulación y resolución de las ecuaciones del H-FGK se torna difícilmente viable. Afortunadamente, y como vimos en la sección \ref{limtrun}, tales ecuaciones admiten una solución independiente del modelo, que se obtiene escogiendo las $H$ como funciones armónicas en el espacio transverso, y con uno de los polos dado por la correspondiente carga, es decir
\begin{equation}
\label{eq:universalsusy}
H^{i} = a^{i}-\frac{p^{i}}{\sqrt{2}}\tau,~~~~r_0=0\, .
\end{equation}

\noindent
Esta elección corresponde a un agujero negro supersimétrico \cite{Tod:1983pm,Behrndt:1997ny, Meessen:2006tu}.

\subsection{La solución supersimétrica general}
\label{sec:susysolution}

Como hemos dicho, introduciendo (\ref{eq:universalsusy}) en (\ref{scalar}) y (\ref{hessiano}) obtenemos una familia de soluciones supersimétricas sin necesidad de resolver ninguna ecuación adicional. La entropía de tal familia vendrá dada por
\begin{equation}
S=\frac{1}{2}\kappa^{0}_{ijk} \frac{p^i p^j p^k}{ \hat{d}_m p^m} W_a\left(s_a \beta\right)\, ,
\end{equation}
\begin{equation}\notag
\beta=\sqrt{\frac{3\hat{n}(\hat{d}_l p^l)^3}{2\kappa^0_{pqr}p^p p^q p^r}}\, ,
\end{equation}

\noindent
mientras que su masa será
\begin{align}
\label{mass}
M =\dot{U}(0)=\frac{1}{2\sqrt{2}}\left[\frac{3\kappa^0_{ijk}p^i a^j a^k}{\kappa^0_{pqr}a^p a^q a^r}\left[1-\frac{1}{1+W_a(s_a\alpha)} \right] -\frac{d_l p^l}{d_n a^n}\left[1-\frac{3}{2\left(1+W_a(s_a\alpha)\right)} \right] \right] \, ,
\end{align}
\begin{equation}
\alpha= \sqrt{\frac{3\hat{n}(d_l a^l)^3}{2\kappa^0_{pqr}a^p a^q a^r}} \, .
\end{equation}

\noindent
En la aproximación que estamos considerando, estamos despreciando términos de orden $\sim e^{-2\pi d_i \Im m z^i}$ con respecto a aquellos yendo como $\sim \pi d_i \Im m z^i e^{-2\pi d_i \Im m z^i}$. Teniendo en cuenta (\ref{scalar}), esta asunción se traduce en la condición
\begin{equation}
W_a(x) e^{-2 W_a(x)}>>e^{-2 W_a(x)}\, .
\end{equation}

\noindent
Está claro que tal condición se satisface para $a=0$ si $x\in [\alpha,\beta]$ para valores positivos suficientemente grandes de $\alpha$ y $\beta$. Sin embargo, no se satisface en absoluto para $x\in [-\frac{1}{e},0)$, que es el rango para el cual ambas ramas de la función de Lambert toman valores reales.

Si asumimos $x \in \left[\alpha,\beta \right]$ para $\alpha,\beta\in\mathbb{R}^+$ suficientemente grandes, $a=0$ y $W_0$ es la única rama real de la función de Lambert. En tal caso, $s=s_0=1$, y tenemos que
\begin{equation}\label{hessian0}
e^{-2U}=\frac{\kappa^{0}_{ijk} H^i H^j H^k}{2\pi \hat{d}_m H^m} W_0\left(\sqrt{\frac{3\hat{n}(\hat{d}_lH^l)^3}{2\kappa^0_{pqr}H^p H^q H^r}}\right)\, ,
\end{equation}
\begin{equation}\label{scalar0}
z^i=i\frac{H^i}{\pi \hat{d}_mH^m}W_0\left( \sqrt{\frac{3\hat{n}(\hat{d}_lH^l)^3}{2\kappa^0_{pqr}H^p H^q H^r}}\right)\, .
\end{equation}

\noindent
En las coordenadas \textit{conformaestáticas} en las que estamos trabajando, es de esperar que el factor de la métrica $e^{-2U}$ diverja en el horizonte de eventos ($\tau\rightarrow -\infty$) como $\tau^2$. Además, hemos de requerir que $e^{-2U}>0$ $\forall \tau \in (-\infty,0]$, e imponer que la solución sea asintóticamente plana $e^{-2U(\tau=0)}=1$. Las dos últimas condiciones son
\begin{equation}\label{posimetri}
\frac{\kappa^0_{ijk} H^i H^j H^k}{2\pi \hat{d}_n H^n}>0 ~~\forall \tau\in (-\infty,0] \, ,
\end{equation}
\begin{equation}\label{asyfla}
 \frac{\kappa^0_{ijk}a^i a^j a^k}{2\pi \hat{d}_m a^m}W_0\left(\alpha \right)=1\, ,
\end{equation}

\noindent
mientras que la primera resulta cumplirse, puesto que
\begin{equation}
e^{-2U}\overset{\tau \rightarrow -\infty}{\longrightarrow} \frac{\kappa^0_{ijk}p^i p^j p^k}{8\pi \hat{d}_m p^m}W_0\left(\beta\right)\tau^2  \,.
\end{equation}

\noindent
(\ref{posimetri}) y (\ref{asyfla}) pueden imponerse en general en cualquier modelo particular que consideremos. Finalmente, la condición para una masa bien definida y positiva puede leerse directamente de (\ref{mass}).

\subsection{Funciones multivaluadas y la conjetura de no-pelo}
Como explicamos en la sección anterior, nuestra aproximación no es consistente con una solución tal que $x\in [-\frac{1}{e},0)$. Esto prohíbe  el dominio en el cual $W(x)$ es una función multivaluada en $\mathbb{R}$ (ambas $W_0$ y $W_{-1}$ toman valores reales en ese intervalo). Sin embargo, parece legítimo preguntarse por las consecuencias que la existencia de una posible solución con dos ramas habría tenido, como habría sido el caso si esta condición no hubiera estado presente. En principio, habríamos podido asignar los límites asintótico ($\tau \rightarrow 0$) y cerca del horizonte ($\tau \rightarrow -\infty$) a cualquier par de valores de los argumentos de $W_0$ y $W_{-1}$ ($x_0$ y $x_{-1}$ respectivamente) a través de una elección adecuada de los parámetros disponibles en la solución. En particular, si hubiéramos elegido $x_{0}|_{\tau=0}=x_{-1}|_{\tau=0}=-1/e$ y $x_{0}|_{\tau \rightarrow -\infty}=x_{-1}|_{\tau \rightarrow -\infty}=\beta$, $\beta \in (-1/e,0)$, ambas soluciones habrían tenido exactamente el mismo límite asintótico (y por tanto los escalares de ambas soluciones habrían coincidido asintóticamente en el infinito espacial), y habríamos estado tratando con dos soluciones regulares completamente distintas con la misma masa\footnote{A pesar de que $W^{\prime}_{0,-1}(x)$ divergen en $x=-1/e$ según se explica en el apéndice \ref{sec:lambert}, y la definición de $M$ involucraría derivadas de la función de Lambert en ese punto, no resultaría difícil curar este comportamiento y conseguir una masa positiva (y finita) imponiendo que $\dot{x}(\tau)\overset{\tau \rightarrow 0 }{\longrightarrow} 0$ más rápido que $|W^{\prime}_{0,-1}(x)|\overset{x \rightarrow -1/e}{\longrightarrow \infty}$.}, cargas y valores de los escalares en infinito, en flagrante contradicción\footnote{Excepto por posibles problemas de estabilidad, que habrían de ser estudiados con la apropiada atención.} con la correspondiente conjetura de unicidad de soluciones de tipo agujero negro y, como consecuencia, con la conjetura de no-pelo.

En este punto, y teniendo en cuenta que nuestra aproximación no es consistente con tal solución bivaluada, la opción de llevar a cabo esta construcción en un contexto diferente al de la teoría de cuerdas no puede ser catalogada de nada más que meramente especulativa. No obstante, una violación de la conjetura de no-pelo en cuatro dimensiones, y más en una teoría \textit{no gaugeada}, tendría consecuencias importantes independientemente de si la solución en cuestión puede ser embebida o no en teoría de cuerdas. En este sentido, la mera posibilidad de que las \textit{ecuaciones de estabilización} puedan admitir (para ciertos prepotenciales, más o menos complicados) soluciones dependientes de funciones multivaluadas parece abrir la puerta hacia posibles violaciones de la conjetura en cuestión en el contexto de la supergravedad $\mathcal{N}=2$ $d=4$. La cuestión (cuya respuesta aceptada es "no", de acuerdo con la comunidad) es ahora: ¿es posible encontrar una teoría de supergravedad \textit{no gaugeada} en cuatro dimensiones con un contenido de campos físicamente aceptable que admita más de una solución estable de tipo agujero negro con la misma masa y cargas eléctricas, magnéticas y escalares? Y si no es así, ¿qué lo impide? Estas cuestiones serán abordadas en una futura publicación \cite{POS}.

\newpage
\section{Conclusiones}
En el presente texto hemos utilizado el formalismo H-FGK (explicado en la sección \ref{HFGK}) para construir nuevas familias de agujeros negros estáticos y esféricamente simétricos, soluciones a la teoría de cuerdas tipo-IIA compactificada en un CY a cuatro dimensiones considerando correcciones perturbativas y no perturbativas en $\alpha^{\prime}$ a la descripción efectiva de la teoría, proporcionada por una teoría de supergravedad $\mathcal{N} = 2$.  

En primer lugar (sección \ref{QBH}) hemos construido una nueva clase de agujeros negros en presencia de la corrección cuántica perturbativa $c$ relacionada con la característica de Euler de la variedad de compactificación. Estos han resultado tener la sorprendente propiedad de carecer de un límite clásico ($c = 0$) bien definido. Tal característica se pone de manifiesto en el hecho de que no solo las soluciones en cuestión dejan de estar bien definidas en ese límite, sino que la misma truncación de campos considerada se torna matemáticamente inconsistente. Esta propiedad sugiere el calificativo de agujeros negros genuinamente cuánticos, y plantea al menos dos cuestiones: ¿cómo de numerosas son, y qué lugar ocupan este tipo de familias en el espacio de soluciones de la tipo-IIA? y ¿es posible llevar a cabo el cálculo de la entropía microscópica para este tipo de soluciones y compararlo con el resultado macroscópico proporcionado por la ley del área de Bekenstein-Hawking?

Paralelamente hemos visto que la regularidad de las soluciones de nuestra familia imponía una condición sobre la estructura topológica del CY de compactificación, y hemos construido explícitamente dos soluciones no extremas para dos modelos correspondientes a sendas variedades de CY con números de Hodge pequeños que satisfacían tal condición. En ambos casos (y también en otros que hemos omitido en este trabajo), los campos constantes resultaban ser constantes, lo que podría indicar que la inclusión de las famosas correcciones cuánticas tiende a estabilizar los campos escalares de la teoría.

Finalmente, en el contexto del modelo \textit{STU}, hemos construido la primera solución de tipo agujero negro no extremo con escalares no constantes en presencia de correcciones perturbativas cuánticas a la tipo-IIA compactificada en un CY a cuatro dimensiones existente en la literatura. 

En la sección \ref{NPBH} hemos considerado los efectos de las correcciones no perturbativas sobre las posibles soluciones, con dos objetivos: por un lado estudiar sus propiedades, interesantes de por sí, y por otro,  determinar si eran capaces de curar las inconsistencias encontradas en nuestras soluciones perturbativas cuando $c = 0$. Hemos comprobado cómo, al menos en el caso supersimétrico (el único que hemos podido resolver en este caso), es posible encontrar soluciones no perturbativas regulares en ausencia de la corrección perturbativa sin que la truncación se vuelva inconsistente. De manera sorprendente, las soluciones construidas vienen dadas en términos de una función multivaluada, lo que da lugar a una posible violación de la conjetura de no-pelo. Hemos argumentado que tal posibilidad no es posible en teoría de cuerdas, pero que permanece abierta en el contexto de la supergravedad no gaugeada cuatridimensional.


\clearpage

\appendix

\section{Álgebras graduadas}
\label{graded}
Un \textit{álgebra $\mathbb{Z}_{N+1}$ graduada} \cite{Shahbazi2} consiste en un espacio vectorial $L$ que es la suma directa de $N+1$ ($N\geq 1$) subespacios $L_k$, i.e.
\begin{equation}
\displaystyle L= \bigoplus_{k=0}^N L_k\, ,
\end{equation}
con un producto $\circ~:~L\times L \longrightarrow~L$, tal que si $u_k\in L_k$ para cada $k$, entonces
\begin{equation}
u_j\circ u_k \in L_{(j+k)~mod~(N+1)}\, .
\end{equation}
A un producto $\circ$ con tal propiedad se le llama \textit{graduación}. En el caso más sencillo, el álgebra graduada consiste en un espacio vectorial que es la suma directa de dos subespacios $L_0$ y $L_1$, $L=L_0\oplus L_1$, y un producto $\circ$ que satisface:
\begin{enumerate}
\item $u_0 \circ v_0 \in L_0$, $\forall u_0,~v_0\in L_0$,
\item $u_0\circ u_1 \in L_1$, $\forall u_0\in L_0,~v_1\in L_1$,
\item $u_1\circ v_1 \in L_0$, $\forall u_1,~v_1\in L_1$,
\end{enumerate}
A este álgebra se le llama \textit{álgebra $\mathbb{Z}_2$-graduada}. Un álgebra de este tipo se convierte en un \textit{álgebra de Lie $\mathbb{Z}_2$-graduada} si su operación satisface, además de la graduación, las propiedades:
\begin{itemize}
\item Supersimetrización: $\forall~x_i\in L_i$, $\forall~x_j\in L_j$ $i,j=0,1$,
$x_i\circ x_j =-(-1)^{ij} x_j \circ x_i$.
\item Identidades de Jacobi generalizadas: $\forall~x_k\in L_k$, $\forall~x_l\in L_l$, $\forall~x_m\in L_m$, $k,l,m=0,1$,
    $x_k\circ (x_l \circ x_m)(-1)^{km}+x_l \circ (x_m \circ x_k)(-1)^{lk}+x_m \circ (x_k\circ x_l)(-1)^{ml}=0$.
\end{itemize}
Con las definiciones de arriba, $L_0$ forma un álgebra de Lie ordinaria por sí mismo, ya que el par $(L_0,\circ)$ satisface las condiciones de tal estructura. Por su parte, el subespacio $L_1$ no forma un álgebra, ya que no es cerrado bajo el producto $\circ$.

A cada elemento homogéneo de un álgebra de lie $\mathbb{Z}_k$-graduada se le asigna por definición un \textit{grado} $g\in\mathbb{Z}_k$ de la siguiente manera: sea $x\in L$, entonces, por definición: $g=g(x)\equiv k\iff x\in L_k$.
Decimos que el elemento $x\in L$ es \textit{par} si $g=0$, y que es \textit{impar} si $g=1$. Con estas definiciones, observamos que el conjunto de generadores $E_i,~i=1,2,3,...,dim~L_0$ que forme una base de $L_0$ estará formada por elementos pares, mientras que los generadores $Q_a,~a=1,...,dim~L_1$ que formen base de $L_1$ serán impares. Al subespacio $L_0$ se le conoce a veces como el \textit{sector bosónico}, mientras que $L_1$ recibe el nombre de \textit{sector fermiónico}.

La definición de grado permite redefinir la operación del álgebra graduada de una forma más compacta según
\begin{equation}
x_{\mu}\circ x_{\nu}\equiv x_{\mu}x_{\nu}-(-1)^{g_{\mu}g_{\nu}}x_{\nu}x_{\mu}=c_{\mu\nu}^{\omega}x_{\omega}\, .
\end{equation}
\noindent
donde $c_{\mu\nu}^{\omega}$ serán las llamadas \textit{constantes de estructura generalizadas} del álgebra graduada,
$c_{\mu\nu}^{\omega}=-(-1)^{g_{\mu}g_{\nu}}c_{\nu\mu}^{\omega}$.

Es sencillo ver ahora que $E_i\circ E_j =[E_i,E_j]$, de forma que la operación entre elementos de $L_0$ es el conmutador, al igual que la operación entre un elemento de $L_0$ y uno de $L_1$: $E_i\circ Q_a =[E_i,Q_a]$. Finalmente, la de dos elementos de $L_1$ es el anticonmutador: $Q_a\circ Q_b =\{Q_a,Q_b\}$.\\

\section{Funciones especiales}
\label{SF}

\subsection{Polilogaritmo}
\label{sec:polylog}
La \textit{función polilogarítmica} o simplemente \textit{polilogaritmo} $Li_w(z)$ (véase por ejemplo \cite{Lewin} para un estudio exhaustivo) es una función especial definida por medio de la serie de potencias
\begin{equation}
Li_w(z)=\sum_{j=1}^{\infty} \frac{z^j}{j^w}\, ,~~z,w\in \mathbb{C}\, .
\end{equation}

\noindent
Esta definición es válida para cualesquiera números complejos $w$ y $z$ para $|z|<1$, pero puede extenderse a $z's$ con $|z|\geq 1$ mediante continuación analítica. De la propia definición es fácil obtener la relación de recurrencia
\begin{equation}
Li_{w-1}(z)=z\frac{\partial Li_w(z)}{\partial z}\, .
\end{equation}
\noindent
El caso $w=1$ corresponde a
\begin{equation}
Li_1(z)=-\log(1-z) \, ,
\end{equation}

\noindent
y de ahí es fácil ver que para $w=-n\in \mathbb{Z}^- \cup \left\{0 \right\}$, el polilogaritmo es una función elemental dada por
\begin{equation}
Li_0(z)=\frac{z}{1-z}\, ,~~ Li_{-n}(z)=\left(z\frac{\partial}{\partial z} \right)^n \frac{z}{1-z}\, .
\end{equation}

\noindent
Los casos especiales $w=2,3$ se denominan \textit{dilogaritmo} y \textit{trilogaritmo} respectivamente, y sus representaciones integrales pueden obtenerse de $Li_1(z)$ haciendo uso de la relación
\begin{equation}
Li_{w}(z)=\int^{z}_0 \frac{Li_{w-1}(s)}{s}ds\, .
\end{equation}

\subsection{Función W de Lambert}
\label{sec:lambert}
La \textit{función W de Lambert} $W(z)$ (también conocida como \textit{logaritmo producto}) lleva el nombre de Johann Heinrich Lambert (1728-1777), quien la introdujo por primera vez en 1758 \cite{Lambert}. Durante sus más de 200 años de historia, ha encontrado numerosas aplicaciones en diferentes áreas de la física (principalmente durante el siglo XX) tales como electrostática, termodinámica (p.e. \cite{Valluri}), física estadística (p.e. \cite{JM}), cromodinámica cuántica (p.e \cite{Gardi:1998qr,Magradze:1998ng,Nesterenko:2003xb,Cvetic:2011vy,Sonoda:2013kia}), cosmología (p.e. \cite{Ashoorioon:2004vm}), mecánica cuántica (p.e. \cite{Mann}) y relatividad general (p.e. \cite{Mann:1996cb}).

$W(z)$ se define implícitamente a través de la ecuación
\begin{equation}
\label{Wf}
z=W(z)e^{W(z)}\, ,~~ \forall z\in \mathbb{C}\,.
\end{equation}

\noindent
Puesto que $f(z)=ze^{z}$ no es un mapa inyectivo, $W(z)$ no está unívocamente definida, y $W(z)$ se refiere genéricamente al conjunto de ramas que resuelven  (\ref{Wf}). Para $W:\mathbb{R}\rightarrow \mathbb{R}$, $W(x)$ tiene dos ramas: $W_0(x_0)$ y $W_{-1}(x_{-1})$, definidas en los intervalos $x_0\in [-1/e,+\infty)$ y $x_{-1}\in [-1/e,0)$ respectivamente (véase la Figura 1). Ambas funciones coinciden en el punto de rama $x=-1/e$, donde $W_0(-1/e)=W_{-1}(-1/e)=-1$. Como consecuencia, la ecuación $x=W(x)e^{W(x)}$ admite dos soluciones reales distintas en el intervalo $x\in [-1/e,0)$.\\
\begin{figure}[h]
 \label{fig:u}
  \centering
    \includegraphics[scale=0.65]{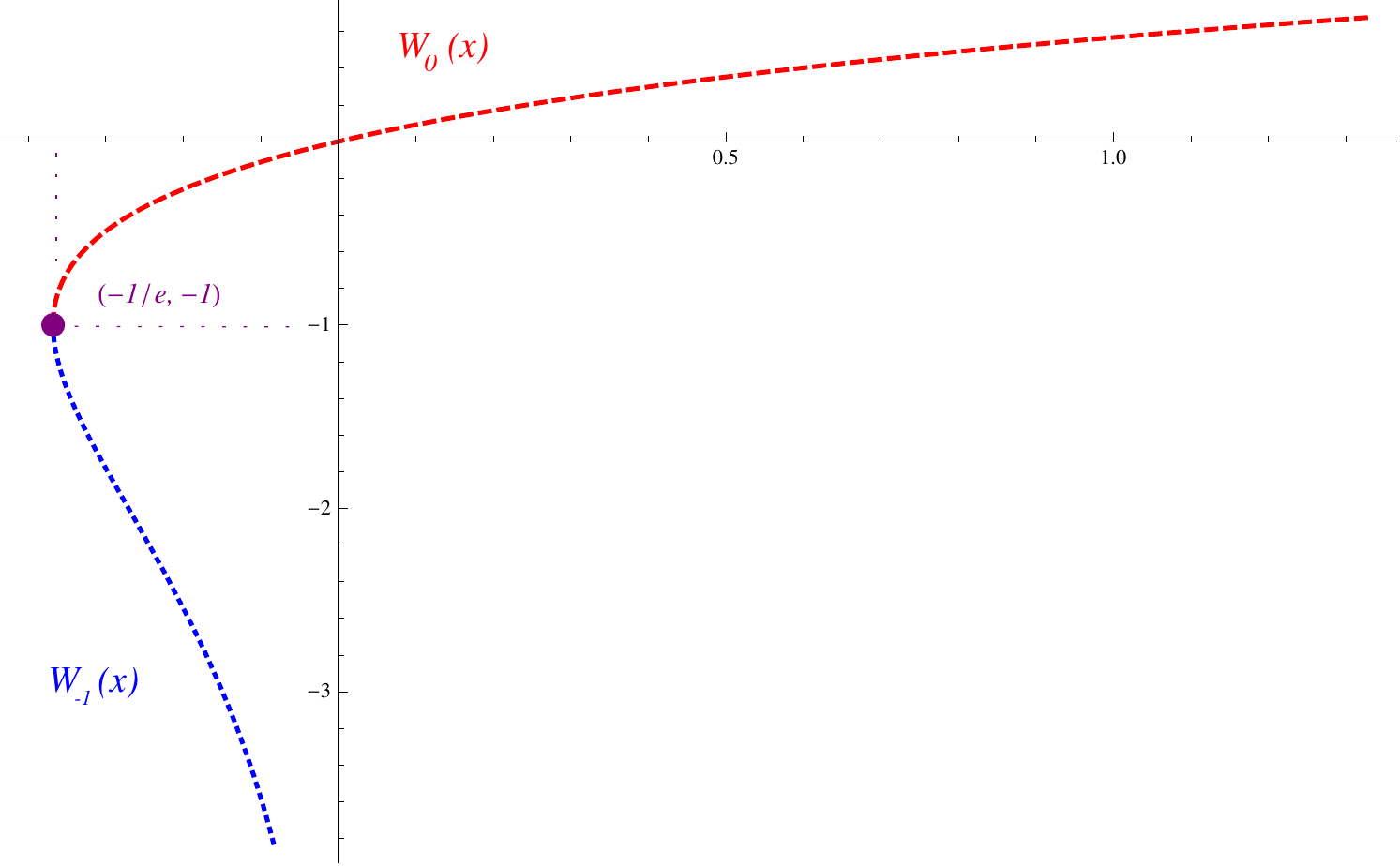}
 \caption{\small{Las dos ramas reales de $W(x)$.}}
\end{figure}

La derivada de $W(z)$ viene dada por
\begin{equation}
\frac{dW(z)}{dz}=\frac{W(z)}{z(1+W(z))},~~\forall z \notin \left\{0,-1/e \right\}, ~~ \frac{dW(z)}{dz}\bigg|_{z=0}=1 \, ,
\end{equation}

\noindent
y no está bien definida para $z=-1/e$ (la función no es diferenciable ahí). En ese punto se tiene
\begin{equation}
\lim_{x\rightarrow -1/e}\frac{dW_0(x)}{dx}=\infty,~~\lim_{x\rightarrow -1/e}\frac{dW_{-1}(x)}{dx}=-\infty \, .
\end{equation}


\renewcommand{\leftmark}{\MakeUppercase{Bibliography}}
\phantomsection
\bibliographystyle{JHEP}
\bibliography{References}
\label{biblio}

\end{document}